%% file: main.tex
\documentclass[sigconf]{acmart}
\usepackage{amsmath}

\usepackage{amssymb}
    \PassOptionsToPackage{numbers, compress}{natbib}

\usepackage[utf8]{inputenc} 
\usepackage[T1]{fontenc}    
\usepackage{hyperref}       
\usepackage{url}            
\usepackage{booktabs}       
\usepackage{amsfonts}       
\usepackage{nicefrac}       
\usepackage{microtype}      
\usepackage{xcolor}         
\definecolor{mygreen}{RGB}{0,150,0}
\usepackage{comment}

\usepackage{mathtools}
\usepackage{amsthm}

\usepackage{multirow}
\usepackage{enumitem}
\usepackage{algorithm}
\usepackage{algorithmic}
\usepackage{makecell} 

\usepackage{bm, bbm}
\usepackage{subcaption}
\usepackage{thm-restate}
\usepackage{wrapfig}
\usepackage{pifont}

\newcount\Comments  
\usepackage{xcolor} 

\title{When Algorithms Mirror Minds: A Confirmation-Aware Social Dynamic Model of Echo Chamber and 
Homogenization Traps}

\author{Ming Tang}
\affiliation{
  \institution{School of Computer Science and Technology, Beijing Jiaotong University}
  \institution{Beijing Key Laboratory of Traffic Data Mining and Embodied Intelligence}
  \institution{Key Laboratory of Big Data \& Artificial Intelligence in Transportation, Ministry of Education}
  \country{China}
}
\email{23120413@bjtu.edu.cn}

\author{Xiaowen Huang}
\authornote{Corresponding Author.}
\affiliation{
  \institution{School of Computer Science and Technology, Beijing Jiaotong University}
  \institution{Beijing Key Laboratory of Traffic Data Mining and Embodied Intelligence}
  \institution{Key Laboratory of Big Data \& Artificial Intelligence in Transportation, Ministry of Education}
  \country{China}
}
\email{xwhuang@bjtu.edu.cn}

\author{Jitao Sang}
\affiliation{
  \institution{School of Computer Science and Technology, Beijing Jiaotong University}
  \institution{Beijing Key Laboratory of Traffic Data Mining and Embodied Intelligence}
  \institution{Key Laboratory of Big Data \& Artificial Intelligence in Transportation, Ministry of Education}
  \country{China}
}
\email{jtsang@bjtu.edu.cn}

\setcopyright{none} 
\renewcommand\footnotetextcopyrightpermission[1]{}

\begin{document}

\begin{abstract}

Recommender systems increasingly suffer from echo chambers and user homogenization, systemic distortions arising from the dynamic interplay between algorithmic recommendations and human behavior. While prior work has studied these phenomena through the lens of algorithmic bias or social network structure, we argue that the psychological mechanisms of users and the closed-loop interaction between users and recommenders are critical yet understudied drivers of these emergent effects. To bridge this gap, we propose the Confirmation-Aware Social Dynamic Model which incorporates user psychology and social relationships to simulate the actual user and recommender interaction process. Our theoretical analysis proves that echo chambers and homogenization traps, defined respectively as reduced recommendation diversity and homogenized user representations, will inevitably occur. We also conduct extensive empirical simulations on two real-world datasets and one synthetic dataset with five well-designed metrics, exploring the root factors influencing the aforementioned phenomena from three level perspectives: the stochasticity and social integration degree of recommender (system-level), the psychological mechanisms of users (user-level), and the dataset scale (platform-level). Furthermore, we demonstrate four practical mitigation strategies that help alleviate echo chambers and user homogenization at the cost of some recommendation accuracy. Our findings provide both theoretical and empirical insights into the emergence and drivers of echo chambers and user homogenization, as well as actionable guidelines for human-centered recommender design. 

\end{abstract}


\maketitle

\input{body}



\bibliographystyle{plainnat}

\bibliography{main}


\appendix

\newpage

\input{appendix}

\end{document}

%% file: body.tex
\section{Introduction}

While recommender systems play a vital role in alleviating information overload across domains such as e-commerce \cite{wang2018billion}, social network \cite{duskin2024echo}, and news platforms \cite{wu2019npa}, they also introduce adverse effects such as echo chambers \cite{nguyen2014exploring, ge2020understanding} and user homogenization \cite{chaney2018algorithmic, mansoury2020feedback} manifested as reduced intra-user content diversity and increased inter-user behavioral similarity, respectively. The former compromises user experience and long-term engagement by limiting exposure to diverse content \cite{gao2023cirs, moller2020not}, while the latter trigger political polarization and other societal issues on social media platforms \cite{tornberg2022digital, garimella2018political}.

These phenomena essentially stem from the interactive feedback loop between users and recommender systems \cite{alatawi2021survey, sun2019debiasing}, in which the system influences users by shaping their exposure and preferences through recommendations, while users in turn affect the system by providing feedback—explicitly or implicitly—that guides future recommendation outputs. Some recent works have simulated the iterative interaction to investigate such phenomena \cite{jiang2019degenerate, kalimeris2021preference, piao2023human, chen2024coevolution, rossi2021closed, chaney2018algorithmic}. However, the aforementioned studies primarily focus on the recommender system itself, while overlooking the role of users in the loop. As noted in \cite{alatawi2021survey}, user-side psychological mechanisms—such as confirmation bias—are equally crucial in the feedback loop and may significantly contribute to the observed negative effects. 

Moreover, real-world recommender systems do not operate in isolation: users interact within social networks, and their preferences can be influenced by their neighbors \cite{chitra2020analyzing, cinus2022effect}. Some opinion dynamics studies \cite{banisch2019opinion, perra2019modelling, chitra2020analyzing, sabin2020pull} have explored the evolution and polarization of user opinions within social networks, but typically without incorporating recommender systems into the dynamic modeling. In contrast, our work jointly examines user–recommender interaction by integrating user psychological mechanisms while also accounting for the role of social networks in real-world scenarios. This comprehensive perspective enables us to investigate the underlying causes behind the emergence and intensification of echo chambers and user homogenization in a more realistic and holistic manner.

The main contributions of this work are as follows:
\begin{itemize}[itemsep=1pt,topsep=0pt,parsep=2pt,partopsep=0pt, leftmargin=1.5em]
\item We propose a novel Confirmation-Aware Social Dynamic Model, which integrates user psychology with social recommendation scenario. On the recommender side, we introduce two \textbf{system-level} parameters to control the degree of recommendation stochasticity and the extent of social information integration. On the user side, we model the strength of users' confirmation and leniency biases during feedback at \textbf{user-level}. This framework enables us to systematically investigate the potential impact of various factors on echo chambers and user homogenization.

\item We establish the equivalent formulation and the sufficient condition for the convergence of our proposed model, and further theoretically demonstrate the inevitability of the reduction in recommendation diversity and the homogenization of user representations. 

\item We simulate varied parameters on two real-world and one synthetic dataset, evaluating echo chambers and user homogenization via five metrics to examine their driving factors. We further adjust the synthetic dataset’s scale (e.g., item/category counts) to analyze how \textbf{platform-level }characteristics influence these effects.

\item We design four mitigation strategies that help alleviate echo chambers and user homogenization, though at the cost of some recommendation accuracy. These findings offer practical insights for designing real-world recommender systems, especially those that prioritize long-term benefits.

\end{itemize}


\section{Confirmation-Aware Social Dynamic Model}
\label{sec:model}
We define a dynamic model incorporating  psychological mechanisms and social propagation to explore the emergence of echo chambers and user homogenization under social recommendation scenario. Consider a setting with $m$ items, $n$ users, and $c$ item categories, first we represent item $j$ according to its associated category :
\(
\mathbf{v}_j = \left[ v_j^{(1)},v_j^{(2)}, \ldots, v_j^{(c)} \right]^T,j \in \{1,2, \ldots,m \}.
\) If item $j$ belongs to $k$ distinct categories \(l_1,l_2, \ldots,l_k\), then \(v_j^{(l)}=\sqrt{1/k}\) if \(l \in \{l_1,l_2, \ldots,l_k \}\), and 0 otherwise. Notably, the vast majority of items are associated with only one category, and let $n_o=\sum_{j=1}^mv_j^{(o)}, o \in \{1,2, \ldots,c \}$ denote the number of items belonging to category $o$. Next, we initialize the representation of user $i$ at time 0 as follows:

\begin{equation}\label{eq:init-u}
\mathbf{u}_i(0) = \frac{\sum_{j \in \mathcal{V}_i^+} \mathbf{v}_j - \sum_{j \in \mathcal{V}_i^-} \mathbf{v}_j}{\| \sum_{j \in \mathcal{V}_i^+} \mathbf{v}_j - \sum_{j \in \mathcal{V}_i^-} \mathbf{v}_j \|_2}. 
\end{equation} 

where $\mathcal{V}_i^+$ and $\mathcal{V}_i^-$ represent the sets of positively and negatively interacted items in the user $i$'s interaction history, respectively. The representation of user $i$ at time $t$ is denoted as \(\mathbf{u}_i(t) = \left[ u_i^{(1)}(t),u_i^{(2)}(t), \ldots, u_i^{(c)}(t) \right]^T,i \in \{1,2, \ldots,n \}\).
\paragraph{(a) A Recommendation Model Integrating Stochasticity and Social integration.}
Some existing works \cite{dean2020designing, kalimeris2021preference, piao2023human, lin2024user} employ the softmax function to map user scores over items, as a relaxation of top-$k$ recommendation, i.e., the probability that user $i$ is recommended item $j$ at time $t$ is given by:

\begin{equation}\label{eq:similar-match}
p^{t}(\mathbf{u}_i, \mathbf{v}_j; \alpha) = \frac{ \exp \left( \alpha \mathbf{v}_j^T \mathbf{u}_i(t) \right) }{ \sum_{k = 1}^m \exp \left( \alpha \mathbf{v}_k^T \mathbf{u}_i(t) \right) }.
\end{equation} 

By sampling $k$ items without replacement from the probability distribution \(
\mathbf{p}_{\mathbf{u}_i}^{t}(\alpha) = \left[ p^{t}(\mathbf{u}_i, \mathbf{v}_1; \alpha), p^{t}(\mathbf{u}_i, \mathbf{v}_2; \alpha),\ldots, p^{t}(\mathbf{u}_i, \mathbf{v}_m; \alpha) \right]
\), we can control the stochasticity of the recommendation list via a temperature parameter $\alpha$. As $\alpha \rightarrow 0$, the distribution approaches uniformity, resulting in purely random recommendations. Conversely, as $\alpha \rightarrow \infty$, the distribution becomes increasingly peaked, amplifying the selection probability of high-scoring items and effectively recovering the classic top-$k$ recommendation. Therefore, \textbf{$\alpha$ serves as a tunable parameter that balances stochasticity and accuracy in recommendation. }

Building upon this recommendation strategy, we incorporate social information and define the social representation of user $i$ as \(
\mathbf{s}_i^{\gamma}(t) = \gamma \mathbf{u}_i(t) + (1 - \gamma) \frac{ \sum_{j \in \mathcal{N}_i} \mathbf{u}_j(t) }{ |\mathcal{N}_i| }
\), where $\mathcal{N}_i$ denotes the set of social neighbors of user $i$, and \textbf{$\gamma$ is a parameter that controls the weight of social information, which means social integration degree}. Consider user dynamics after incorporating social information, the probability that user $u$ is recommended item $j$ at time $t$ is given by:

\begin{equation}\label{eq:social-similar-match}
p^{t}(\mathbf{s}_i^{\gamma}, \mathbf{v}_j; \alpha) = \frac{ \exp \left( \alpha \mathbf{v}_j^T \mathbf{s}_i^{\gamma}(t) \right) }{ \sum_{k = 1}^m \exp \left( \alpha \mathbf{v}_k^T \mathbf{s}_i^{\gamma}(t) \right) }.
\end{equation} 

when $\gamma=1$, this function degenerates to Equation \ref{eq:similar-match}. We then sample $h$ items without replacement from the probability distribution \(
\mathbf{p}_{\mathbf{s}_i}^{t}(\alpha) = \left[ p^{t}(\mathbf{s}_i^{\gamma}, \mathbf{v}_1; \alpha), p^{t}(\mathbf{s}_i^{\gamma}, \mathbf{v}_2; \alpha),\ldots, p^{t}(\mathbf{s}_i^{\gamma}, \mathbf{v}_m; \alpha) \right]
\) to construct the recommendation list for user $i$, denoted as \(\mathcal{R}_i^t=\{r_{i}^1, r_{i}^2, \ldots, r_{i}^h\}\).

\paragraph{(b) User Feedback Mechanism under Confirmation and Leniency biases.} 

\begin{figure}[t]
    \centering
    \includegraphics[width=0.5\textwidth]{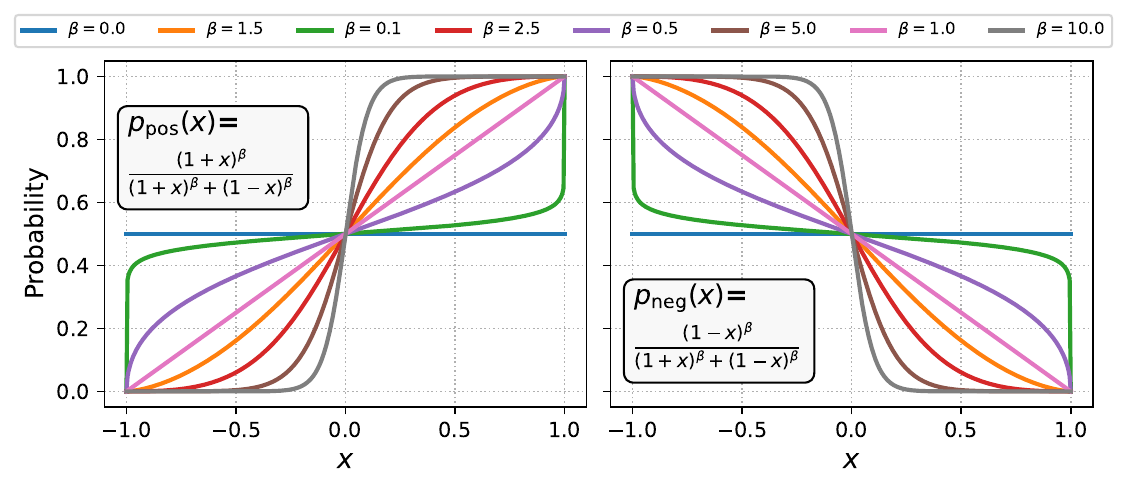}
    \vspace{-10pt}
    \caption{Positive and negative feedback probabilities under varying $\beta$.}
    \label{fig:confirmation-bias}
\end{figure}

Confirmation bias refers to the user's cognitive tendency to prefer information that aligns with their existing preferences while disregarding or avoiding dissimilar content \cite{alatawi2021survey, rieger2021item}, and we investigate the impact of confirmation bias strength on the dynamics of user–recommender interactions. Specifically, after recommending item $j$ to user $i$ at time $t$, we simulate the user's positive or negative feedback with confirmation bias through a carefully designed function, as shown in Figure \ref{fig:confirmation-bias}. The probabilities of positive and negative feedback are \(p_{pos}^t(\mathbf{u}_i, \mathbf{v}_j; \beta, \epsilon)\) and \(p_{neg}^t(\mathbf{u}_i, \mathbf{v}_j; \beta, \epsilon)\), respectively.
\begin{equation}
p_{pos}^t(\mathbf{u}_i, \mathbf{v}_j; \beta, \epsilon)= \frac{ \left( 1 + \mathbf{v}_j^T \mathbf{u}_i(t) \right)^\beta  }{ \left( 1 + \mathbf{v}_j^T \mathbf{u}_i(t) \right)^\beta + \left( 1 - \mathbf{v}_j^T \mathbf{u}_i(t) \right)^\beta } + \frac{\epsilon}{2}
\end{equation}
\begin{equation}
p_{neg}^t(\mathbf{u}_i, \mathbf{v}_j; \beta, \epsilon)= \frac{ \left( 1 - \mathbf{v}_j^T \mathbf{u}_i(t) \right)^\beta  }{ \left( 1 + \mathbf{v}_j^T \mathbf{u}_i(t) \right)^\beta + \left( 1 - \mathbf{v}_j^T \mathbf{u}_i(t) \right)^\beta } - \frac{\epsilon}{2}.
\end{equation}
Formally, when the user-item score \(\mathbf{v}_j^T \mathbf{u}_i(t) >0\), increasing the confirmation bias parameter 
\(\beta\) monotonically raises the likelihood of positive feedback \(p_{pos}\) while suppressing negative feedback \(p_{neg}\), demonstrating users' tendency to amplify preference-aligned content. The inverse relationship holds for \(\mathbf{v}_j^T \mathbf{u}_i(t) < 0\), where larger \(\beta\) exacerbates rejection of disagreeable information. \textbf{\(\beta\) serves as a tunable parameter that controlling confirmation bias intensity in user's feedback psychology}. 


The parameter $\epsilon$ models leniency bias~\cite{seaward2023rating} in user feedback psychology, reflecting a tendency to give biased responses (positive/negative) irrespective of item-user preference alignment. \textbf{As a tunable parameter,  $\epsilon$ controls leniency bias strength}: higher values indicate positive feedback tendencies, while lower values indicate negative tendencies.

\paragraph{(c) Dynamic User Representation Update.}
After user $i$ receives the recommendation list \(\mathcal{R}_i^t=\{r_{i}^1, r_{i}^2, \ldots, r_{i}^h\}\) and provides positive or negative feedback, we update the user representation as follows, with $\eta$ controlling the update rate:

\begin{equation}\label{eq:user-update}
\mathbf{u}_i(t+1) = \mathbf{u}_i(t) + \frac{\eta}{h} \sum_{j=1}^h w_i^t(r_i^j) \mathbf{v}_{r_i^j};
\end{equation}
\begin{equation}
where\ w_i^t(r_i^j) =
\begin{cases}
1, & \text{with probability } p_{pos}^t(\mathbf{u}_i, \mathbf{v}_{r_i^j}; \beta, \epsilon) \\
-1, & \text{with probability } p_{neg}^t(\mathbf{u}_i, \mathbf{v}_{r_i^j}; \beta, \epsilon)
\end{cases}\nonumber
\end{equation}

\section{Theoretical Analysis and Derivations}
After defining our dynamic model, we first prove its equivalent mathematical formulation and a sufficient condition for convergence, demonstrating that the model admits convergent cases. We then show that, as the number of interaction rounds increases, user homogenization and recommendation diversity decline are inevitable outcomes. We provide the detailed proof in Appendix \ref{apd:thm1}, \ref{apd:thm2}, \ref{apd:thm3}, \ref{apd:thm4}.
\begin{proposition}\label{thm:mat-format} (Equivalence Characterization)

\begin{itemize}[noitemsep,topsep=0pt,parsep=2pt,partopsep=0pt, leftmargin=1.5em]
\item Equation \ref{eq:user-update} is equivalent to the following form: 
\begin{equation}\label{eq:user-update-equiv}
\mathbf{u}_i(t+1) =  \mathbf{u}_i(t) + \eta \sum_{j=1}^mp^{t}(\mathbf{s}_i^{\gamma}, \mathbf{v}_j; \alpha) g^{t}( \mathbf{u}_i, \mathbf{v}_j; \beta, \epsilon ) \mathbf{v}_{j}; 
\end{equation}
where \(
g^{t}( \mathbf{u}_i, \mathbf{v}_j; \beta, \epsilon )=p_{pos}^t(\mathbf{u}_i, \mathbf{v}_{r_i^j}; \beta, \epsilon)-p_{neg}^t(\mathbf{u}_i, \mathbf{v}_{r_i^j}; \beta, \epsilon)
\).
\item Let social matrix \(\mathbf{S}=[S_{ij}]_{i,j=1}^n \in \mathbb{R}^{n \times n}\), where \(S_{ij} = 1\) if user $i$ trusts user $j$ or a social connection exists between them, and \(S_{ij} = 0\) otherwise; and let user matrix \(\mathbf{U}(t)=[\mathbf{u}_1(t), \mathbf{u}_2(t), \ldots, \mathbf{u}_n(t)] \in \mathbb{R}^{c \times n}\), item matrix \(\mathbf{V}=[\mathbf{v}_1, \mathbf{v}_2, \ldots, \mathbf{v}_m] \in \mathbb{R}^{c \times m}\). Under these definitions, Equation \ref{eq:user-update-equiv} admits the equivalent matrix equation:
\begin{equation}\label{eq:update-mat-equiv}
\mathbf{U}(t+1) = \mathbf{X} + \mathbf{Y} \mathbf{U}(t) + \mathbf{Z} \mathbf{U}(t) \mathbf{\tilde{S}}^T;
\end{equation}
where \(\mathbf{X}=\frac{\eta \epsilon}{m}\mathbf{V}\mathbf{1}_{m \times n} \in \mathbb{R}^{c \times n}\ ;\\\mathbf{Y}=\mathbf{I}_c+\frac{\eta(\alpha \epsilon \gamma+\beta)}{m}\mathbf{V}\mathbf{V}^T-\frac{\eta \alpha \epsilon \gamma}{m^2}\mathbf{V}\mathbf{1}_{m \times m}\mathbf{V}^T\in \mathbb{R}^{c \times c}\ ;\\\mathbf{Z}=\frac{\eta\alpha \epsilon (1-\gamma)}{m}\mathbf{V}\mathbf{V}^T-\frac{\eta\alpha \epsilon (1-\gamma)}{m^2}\mathbf{V}\mathbf{1}_{m \times m}\mathbf{V}^T \in \mathbb{R}^{c \times c}\ ;\\\mathbf{\tilde{S}}=\operatorname{diag}(\mathbf{S}\mathbf{1}_{n \times 1})^{-1}\mathbf{S}\in \mathbb{R}^{n \times n}\ ;\ \mathbf{1}_{m \times n} \in \mathbb{R}^{m \times n}\)\ denotes the all-ones matrix\(\ ;\  \mathbf{I}_c \in \mathbb{R}^{c \times c}\) denotes the identity matrix with ones on the diagonal and zeros elsewhere.
\end{itemize}
\end{proposition}

\begin{proposition}\label{thm:convergence} (Convergence Condition)

The sufficient condition for the convergence of Equation \ref{eq:update-mat-equiv} is given by:
\begin{equation}
\eta(\frac{\beta}{2}+\frac{\alpha\epsilon\gamma}{2}+\frac{\beta^2}{8\alpha\epsilon\gamma}) < 1    
\end{equation}
Under this condition, the Equation \ref{eq:update-mat-equiv} converges to the closed-form solution:
\begin{equation}
\mathbf{U}^* = \operatorname{unvec}((\mathbf{I}_{nc}-(\mathbf{I}_n \otimes \mathbf{Y}+\mathbf{\tilde{S}} \otimes \mathbf{Z}))^{-1}\operatorname{vec}(\mathbf{X}))  
\end{equation} where $\otimes$ denotes the Kronecker product, \(\operatorname{vec}(\cdot):\mathbb{R}^{c \times n} \rightarrow \mathbb{R}^{cn}\) denotes the column-wise vectorization operator, \(\operatorname{unvec}(\cdot):\mathbb{R}^{cn} \rightarrow \mathbb{R}^{c \times n}\) denotes the inverse vectorization operation.

\end{proposition}

\begin{corollary}\label{thm:cluster} (User Homogenization)

Consider user dynamics with parameters 
\(
\epsilon = 0, \gamma=1
\), if there exists a feature dimension $k$ satisfying for any user pair $(i,j)$ at time $t$:
\begin{equation}
u_i^{(k)}(t)u_j^{(k)}(t) \geq \frac{\|\mathbf{u}_i(t)\|_2\|\mathbf{u}_j(t)\|_2}{\sqrt{1+(\frac{2n_k+\lambda n_k^2}{\sum_{l \neq k}2n_l+\lambda n_l^2})^2}}
\end{equation}
where $\lambda = \eta \beta/m$, then the following conclusions hold:
\begin{itemize}[noitemsep,topsep=0pt,parsep=2pt,partopsep=0pt, leftmargin=1.5em]
\item Transient Homogenization
\[
\mathbf{u}_i^T(t+1)\mathbf{u}_j(t+1) \geq \mathbf{u}_i^T(t)\mathbf{u}_j(t)
\]
\item Steady-state Homogenization

if \(k = \operatorname*{argmax}_{o \in \{1,2, \ldots,c \}} \, n_o\), then \(\forall \tau \in \mathbb{Z}^{+}, \mathbf{u}_i^T(t+\tau)\mathbf{u}_j(t+\tau) \geq \mathbf{u}_i^T(t+\tau-1)\mathbf{u}_j(t+\tau-1)\)
\end{itemize}

\end{corollary}

\begin{corollary}\label{thm:entropy} (Entropy Decay)

Let \(s_i^{o}(t)=\left| \{ r \in \mathcal{R}_i^t \mid r \in Category \ o \} \right|\) denote the count of items from category $o$ in the user $i$'s recommendation list $\mathcal{R}_i^t$ at time $t$, then the entropy of recommended categories is defined as:
\begin{equation}\label{eq:entropy-defin}
H_i(t)=-\sum_{o=1}^c\frac{s_i^o(t)}{\sum_{k=1}^cs_i^k(t)}\ln \frac{s_i^o(t)}{\sum_{k=1}^cs_i^k(t)}
\end{equation}
Furthermore, we theoretically demonstrate that $H_i(t)$ decreases over time with parameters 
\(
\epsilon = 0, \gamma=1
\), reflecting a gradual loss of diversity in the recommendation lists.
\end{corollary}

\section{Experiments}
\label{sec:experiments}

\begin{table*}[t]
\vspace{-15pt}
  \caption{Statistics of datasets. }
  \label{table:dataset}
  \centering
  \scalebox{0.82}{
  \begin{tabular}{c|c|ccccccc}
    \toprule
    \multicolumn{2}{c|}{\textbf{Dataset}}  &  Users  &  Items  & Categories & Interactions & Social Links  &  Social Density  \\
    \midrule
    \multirow{2}{*}{\makecell[c]{Real-world}} 
    & Ciao &  7,375  &  105,114  & 28  &  284,086 & 111,781  &  0.206\% \\
    
    & Epinions & 22,164 & 296,277 & 27 & 922,267 & 300,548 & 0.061\%\\
    \midrule
    \multirow{4}{*}{\makecell[c]{Synthetic}} 
    & Normal &  1,000  &  10,000  & 10 & / & 10,000  &  1.000\% \\
    & Varied $m$ &  1,000  &  [500,\ 100,000]  & 10 & / & 10,000  &  1.000\% \\
    & Varied $\mathbf{|S|}$ &  1,000  &  10,000  & 10 & / & [500,\ 200,000]  &  [0.05\%,\ 20.0\%] \\
    & Varied $c$ &  1,000  &  10,000  & [5,\ 14] & / & 10,000  &  1.000\% \\
    \bottomrule
  \end{tabular}
  }
  
  \vspace{-5pt}
\end{table*}


\subsection{Datasets and Implementation.}
We conduct extensive experiments on two real-world datasets, Ciao \cite{tang2012mtrust} and Epinions \cite{tang2012etrust}, along with a synthetic dataset that enables controlled manipulation of dataset scale for deeper analysis of its influence on echo chambers and user homogenization. 

In the real-world datasets, when initializing user interest representations in Equation \ref{eq:init-u}, we treat the set of items with ratings greater than or equal to 3 in the interaction history as $\mathcal{V}^+$, and those with ratings less than 3 as $\mathcal{V}^-$. 

In the synthetic dataset, we randomly initialize users’ interest representations with $\ell_2$-normalization. For items, a category is randomly assigned to each item, after which its representation is generated following the format specified in the beginning of Section \ref{sec:model}. Social relationships are constructed by randomly generating edges, given a fixed total number of links. 

Table \ref{table:dataset} summarizes the details of the datasets used in our experiments. And the implementation details are provided in Appendix \ref{apd:implementation}.

\subsection{Metrics.}
To quantify the echo chambers and user homogenization effects at time $t$, we propose two individual-level metrics (\(\mathbf{RCE(t)}\) and \(\mathbf{RA(t)}\)) and three group-level metrics (\(\mathbf{ND(t)}\), \(\mathbf{PDV(t)}\) and \(\mathbf{TS@k(t)}\)), all derived from the $\ell_2$-normalized user representation $\mathbf{u}_i(t), i \in \{1, 2, \ldots, n\}$ and the corresponding recommendation list \\
\(\mathcal{R}_i^t=\{r_{i}^1, r_{i}^2, \ldots, r_{i}^h\}\).
\begin{itemize}[itemsep=1pt,topsep=0pt,parsep=2pt,partopsep=0pt, leftmargin=1.5em]

\item \emph{Recommendation Category Entropy (RCE)}: We define the average category entropy of users' 
recommendation lists at time \(t\) as \(\mathbf{RCE(t)} = \frac{1}{n} \sum_{i=1}^{n} H_i(t)\), where \(H_i(t)\) denotes the category 
entropy of user \(i\)'s recommendation list, as defined in Equation~\ref{eq:entropy-defin}. A smaller \(\mathbf{RCE(t)}\) indicates lower diversity in recommended item categories.

\item \emph{Recommendation Accuracy (RA)}: We define the proportion of recommended items that align with user interests at time \(t\) as
\(\mathbf{RA(t)} = \frac{1}{nh} \sum_{i=1}^n \sum_{k=1}^{h} \mathbb{I}\left( \mathbf{u}_i(t)^T \mathbf{v}_{r_i^k} > 0.7 \right)\),
where \(\mathbb{I}(\cdot)\) denotes the indicator function that returns 1 if the condition holds and 0 otherwise. A larger \(\mathbf{RA(t)}\) indicates better alignment between recommended items and users' preferences, reflecting a higher recommendation accuracy rate.

\item \emph{Neighbor Distance (ND)}:We define the average distance between socially connected users at time \(t\) as
\\
\(
\mathbf{
ND(t)} = \frac{1}{|\mathcal{E}_S|} \sum_{(i,j) \in \mathcal{E}_S} \left\| \mathbf{u}_i(t) - \mathbf{u}_j(t) \right\|_2,
\)
where \(\mathcal{E}_S = \{(i,j) \mid S_{ij} = 1\}\) denotes the set of user pairs with a social connection. A smaller \(\mathbf{ND(t)}\) indicates a higher degree of social homogenization.

\item \emph{Pairwise Distance Variance (PDV)}: We define the variance of pairwise distances among all distinct users at time \(t\) as
\(
\mathbf{PDV(t)} = \frac{2}{n(n-1)} \sum_{i=1}^{n-1} \sum_{j=i+1}^{n} \left( \left\| \mathbf{u}_i(t) - \mathbf{u}_j(t) \right\|_2 - \mu(t) \right)^2,
\)
where \(\mu(t) = \frac{2}{n(n-1)} \sum_{i=1}^{n-1} \sum_{j=i+1}^{n} \left\| \mathbf{u}_i(t) - \mathbf{u}_j(t) \right\|_2\) denotes the mean pairwise distance. A larger \(\mathbf{PDV(t)}\) indicates stronger user polarization and local clustering.

\item \emph{Top-k-Similarity (TS@k)}: We define the average inner product between each user and their top-\(k\) most similar users at time \(t\) as
\(
\mathbf{TS@k(t)} = \frac{1}{n k} \sum_{i=1}^{n} \sum_{j \in \mathcal{N}_i^k(t)} \mathbf{u}_i(t)^T \mathbf{u}_j(t),
\)
where \(\mathcal{N}_i^k(t)\) denotes the set of top-\(k\) users most similar to user \(i\) based on the inner product of user representations. A larger \(\mathbf{TS@k(t)}\) indicates a greater degree of local user homogenization.

\end{itemize}

\subsection{Phenomena and Findings}

\subsubsection{Effect of Key Parameters $\alpha$, $\beta$, $\gamma$ and $\epsilon$ in the Ciao Dataset.}

\begin{figure*}[h]
  \centering
  \includegraphics[width=0.8\linewidth]{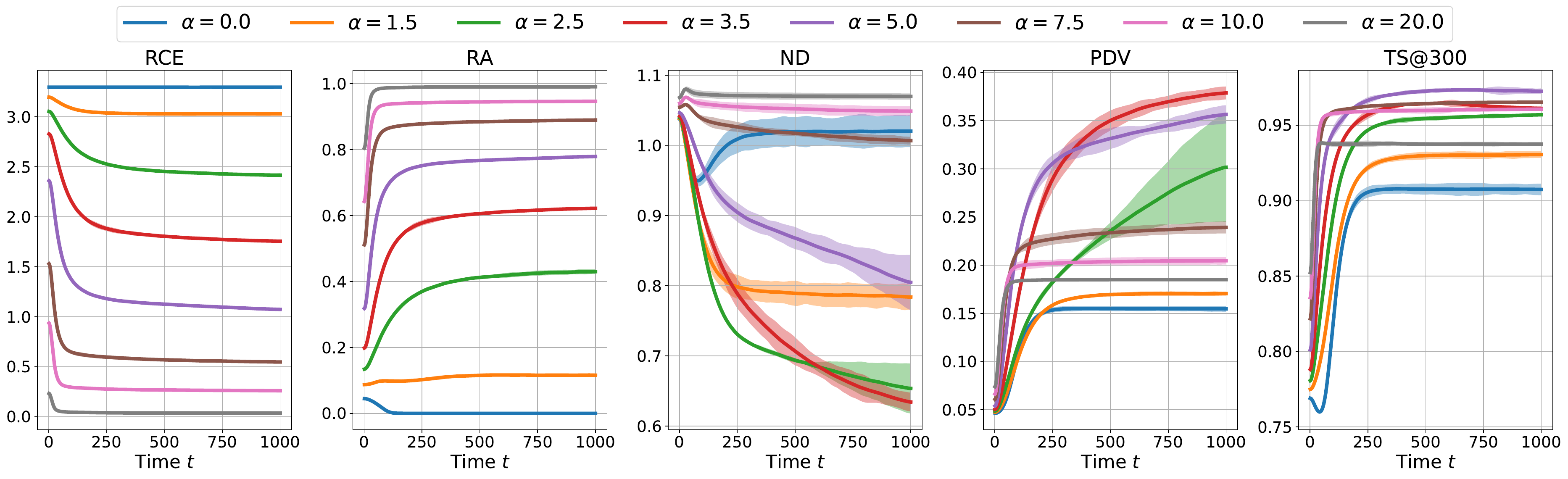}
  \caption{Metric trends over time under varying $\alpha$ on Ciao dataset.}
  \label{fig:ciao_alpha}
\end{figure*}

\paragraph{\textbf{Sensitivity Effect $\boldsymbol{\alpha}$.}}
As mentioned in Section \ref{sec:model}(a), a larger value of $\alpha$ indicates a more personalized recommendation algorithm. When $\alpha \rightarrow \infty$, the situation becomes equivalent to standard top-$k$ recommendation, whereas $\alpha = 0$ corresponds to random recommendation. 

Figure \ref{fig:ciao_alpha} illustrates the temporal dynamics of all metrics on the Ciao dataset under different values of $\alpha$, and the following observations can be drawn: 1) For all values of $\alpha$, RCE consistently decreases while RA steadily increases over time, reflecting the echo chamber phenomenon and the inherent trade-off between recommendation diversity and accuracy. Furthermore, as $\alpha$ increases and the recommendation becomes more personalized, the overall level of RCE declines while RA increases, reinforcing this trade-off. These trends with narrow confidence intervals suggest high stability and consistency across different experimental runs. 2) Within each setting of $\alpha$, ND generally decreases over time, indicating that socially connected users become more similar as their preferences are learned. However, when comparing across different $\alpha$ values, the final ND exhibits a non-monotonic trend—first decreasing and then increasing. This suggests that moderate personalization reduces the differences between socially connected users, while excessive personalization leads to more polarized and individualized user representations. 3) At each level of $\alpha$, both PDV and TS exhibit increasing trends over time, indicating that the iterative interaction between users and the recommender system tends to reinforce overall user polarization and local homogenization. However, when comparing across different values of $\alpha$, we observe that the overall levels of PDV and TS first increase and then decrease as $\alpha$ grows. A possible explanation is that moderate personalization accelerates the convergence of user representations toward a few dominant interest patterns, which intensifies both global polarization and local similarity. While personalization becomes overly strong, each user is steered toward highly specific and individualized content. While this may push individual user representations toward more extreme points in the representation space, it also disperses them more sparsely across the space, thereby slightly alleviating the concentration and overlap that lead to strong homogenization and polarization.

\begin{figure*}[h]
  \centering
  \includegraphics[width=0.8\linewidth]{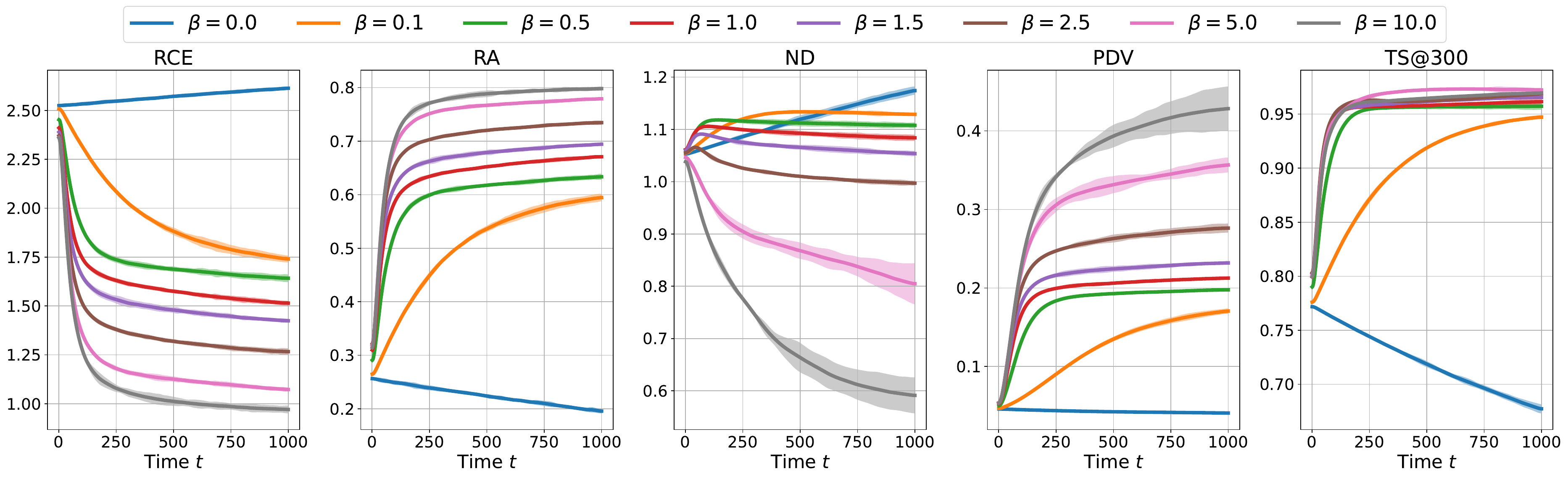}
  \caption{Metric trends over time under varying $\beta$ on Ciao dataset.}
  \label{fig:ciao_beta}
\end{figure*}
\paragraph{\textbf{Sensitivity Effect $\boldsymbol{\beta}$.}}
As mentioned in Section \ref{sec:model}(b), a larger value of $\beta$ reflects stronger confirmation bias, which drives users to respond more extremely—reinforcing agreement with preference-aligned content and intensifying rejection of incongruent information. When $\beta = 0$, users ignore their own preferences and provide positive or negative feedback entirely at random.

Figure \ref{fig:ciao_beta} illustrates the temporal dynamics of all metrics on the Ciao dataset under different values of $\beta$, and the following observations can be drawn: 1) Similar to the results observed under varying 
$\alpha$, the trends of RCE and RA under different values of $\beta$ reflect a trade-off between recommendation diversity and accuracy. As $\beta$ increases, diversity further decreases as the echo chambers intensifies, while accuracy  continues to improve. 2) As $\beta$ increases, the trend of ND shifts from rising to falling, with the decline becoming increasingly pronounced. This suggests that stronger confirmation bias (i.e., more extreme user feedback) intensifies social homogenization. 3) PDV and TS exhibit an overall increasing trend; however, as $\beta$ increases, PDV significantly amplifies this upward trend, while TS shows a moderate enhancement. This indicates that stronger confirmation bias directly intensifies both global polarization and local homogenization.

\begin{figure*}[h]
  \centering
  \includegraphics[width=0.8\linewidth]{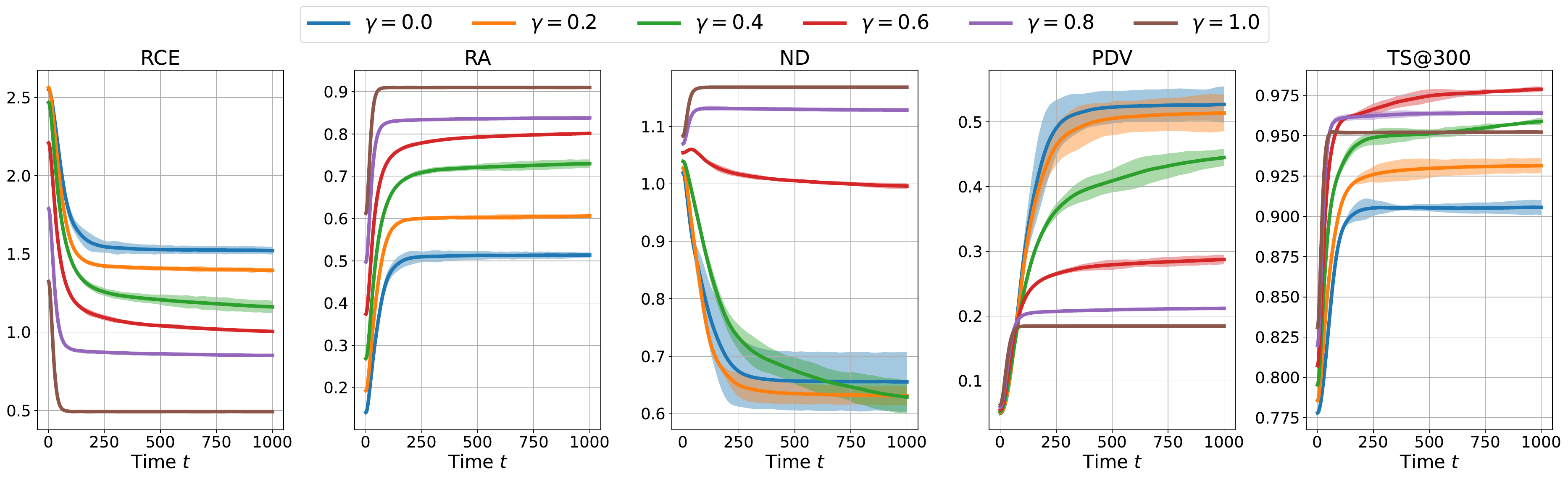}
  \caption{Metric trends over time under varying $\gamma$ on Ciao dataset.}
  \label{fig:ciao_gamma}
\end{figure*}
\paragraph{\textbf{Sensitivity Effect $\boldsymbol{\gamma}$.}}
As mentioned in Section \ref{sec:model}(a), a smaller value of $\gamma$ indicates a greater incorporation of social information during recommendation. Specifically, $\gamma$=0 means the recommendation relies solely on the interest representations of social neighbors, whereas $\gamma$=1 corresponds to recommendations based entirely on the user's own interest representation.

Figure \ref{fig:ciao_gamma} illustrates the temporal dynamics of all metrics on the Ciao dataset under different values of $\gamma$, and the following observations can be drawn: 1) Similar to the results observed under varying 
$\alpha$ and $\beta$, the trends of RCE and RA under different values of $\gamma$ reflect a trade-off between recommendation diversity and accuracy. A larger value of $\gamma$ indicates a higher weight placed on the user's own interests, resulting in more personalized recommendations. 2) On the ND curves, decreasing $\gamma$ (which is equivalent in effect to increasing $\beta$) clearly shows that incorporating more social information into the recommendation process leads to greater social homogenization. 3) Although both PDV and TS exhibit an overall increasing trend, a larger $\gamma$ suppresses the growth of PDV, indicating that incorporating more social information during recommendation intensifies global polarization—likely due to community clustering effects inherent in the social network structure. In contrast, increasing $\gamma$ first amplifies the growth of TS and then mitigates it, suggesting that a moderate level of social information integration leads to more pronounced local homogenization.

\begin{figure*}[h]
  \centering
  \includegraphics[width=0.8\linewidth]{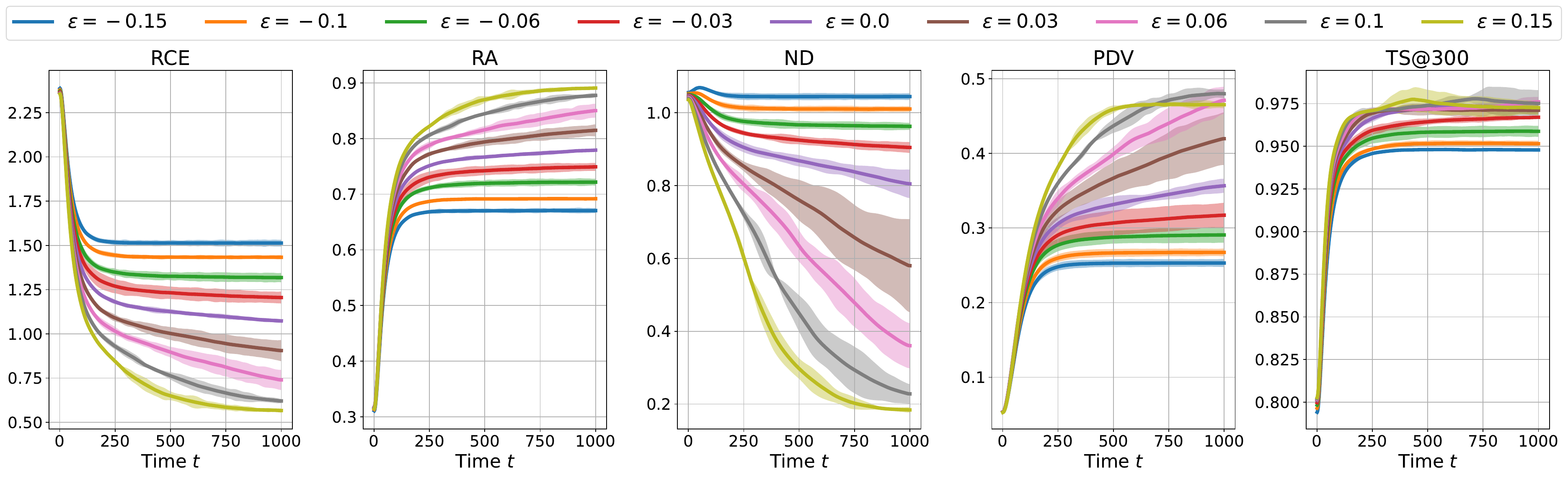}
  \caption{Metric trends over time under varying $\epsilon$ on Ciao dataset.}
  \label{fig:ciao_epsilon}
\end{figure*}

\begin{figure*}[!h]
  \centering
  \includegraphics[width=0.8\linewidth]{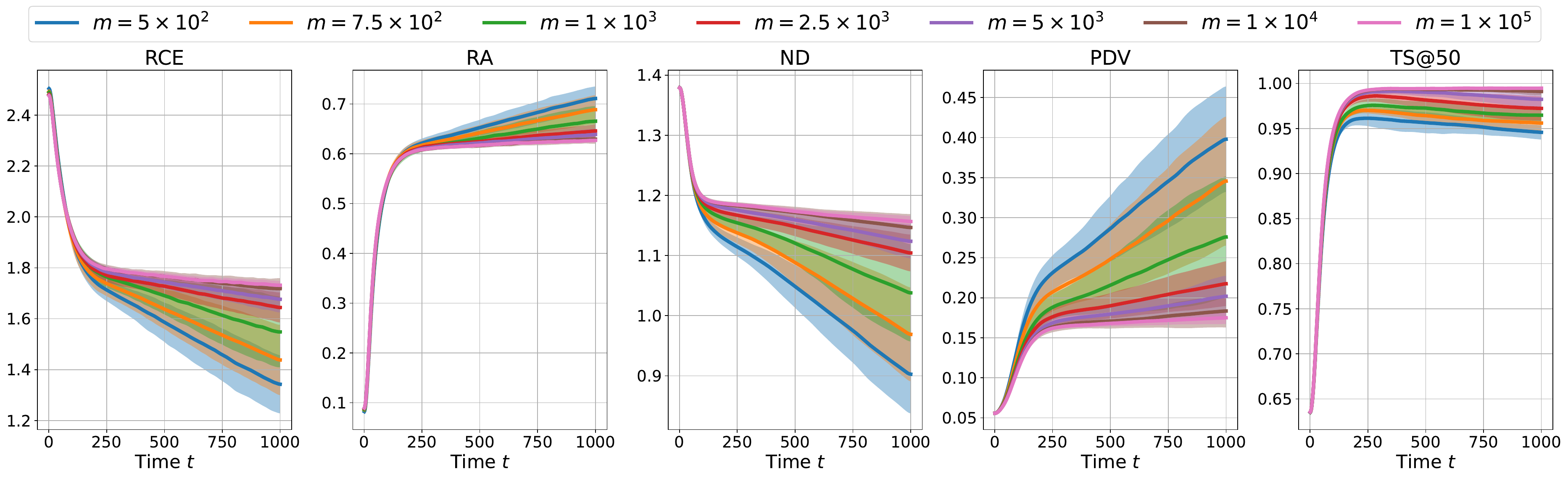}
  \caption{Metric trends over time under varying $m$ on Synthetic dataset.}
  \label{fig:synthetic_m}
\end{figure*}
\begin{figure*}[!h]
  \centering
  \includegraphics[width=0.8\linewidth]{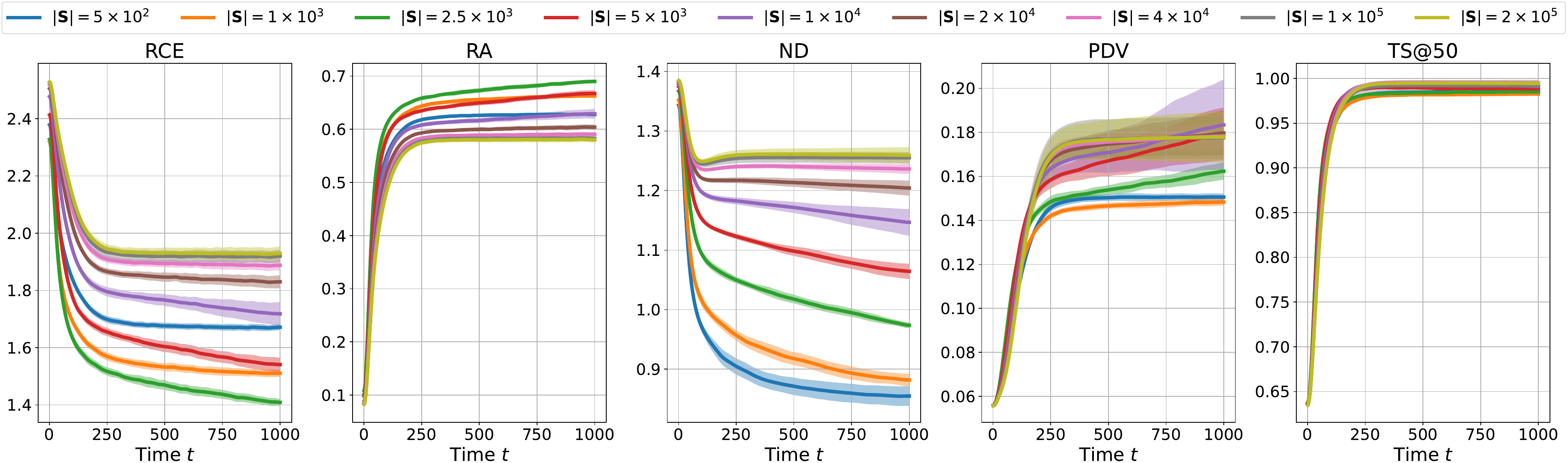}
  \caption{Metric trends over time under varying $|\mathbf{S}|$ on Synthetic dataset.}
  \label{fig:synthetic_s}
\end{figure*}
\paragraph{\textbf{Sensitivity Effect $\boldsymbol{\epsilon}$.}}
As mentioned in Section \ref{sec:model}(b), a larger value of $\epsilon$ indicates stronger leniency bias that users tend to provide positive feedback regardless of their actual interest.
Figure \ref{fig:ciao_epsilon} illustrates the temporal dynamics of all metrics on the Ciao dataset under different values of
$\epsilon$, and the following observations as 
$\epsilon$ increases can be drawn: 1)  Recommendation diversity decreases as the echo chambers intensifies, and users receive more items that align with their preferences. 2) The average distance between social neighbors (ND) shrinks, indicating increased social homogenization. 3) Both PDV and TS rise, signaling stronger global polarization and local homogenization.

\subsubsection{Effect of Dataset Scale $m$, $|\mathbf{S}|$, and $c$ in the Synthetic Dataset}

\paragraph{\textbf{Sensitivity Effect $\boldsymbol{m}$.}}
As the number of items $m$ increases in Figure \ref{fig:synthetic_m}: 1) Recommendation diversity (RCE) increases while relevance (RA) decreases. 2) Both the social neighbor distance (ND) and user similarity (TS) increase.  3) The polarization degree (PDV) decreases. These results suggest that increasing $m$ can mitigate echo chambers, social homogenization, and global polarization by providing users with access to a broader range of content. This richer item pool helps diversify recommendation outcomes and reduces overlap among users’ preferences, but at the cost of intensified localized homogenization.

\paragraph{\textbf{Sensitivity Effect $\boldsymbol{\mathbf{|S|}}$.}}
As the number of social links $\mathbf{|S|}$ increases in Figure \ref{fig:synthetic_s}: 1) RCE first decreases and then increases, indicating that a moderately dense social network may lead to severe echo chambers. 2) Both ND and PDV increase. These results suggest that higher social network density mitigates social homogenization by helping users differentiate from their neighbors, but also drives global polarization by reinforcing personalized extremes.

\begin{figure*}[!h]
  \centering
  \includegraphics[width=0.8\linewidth]{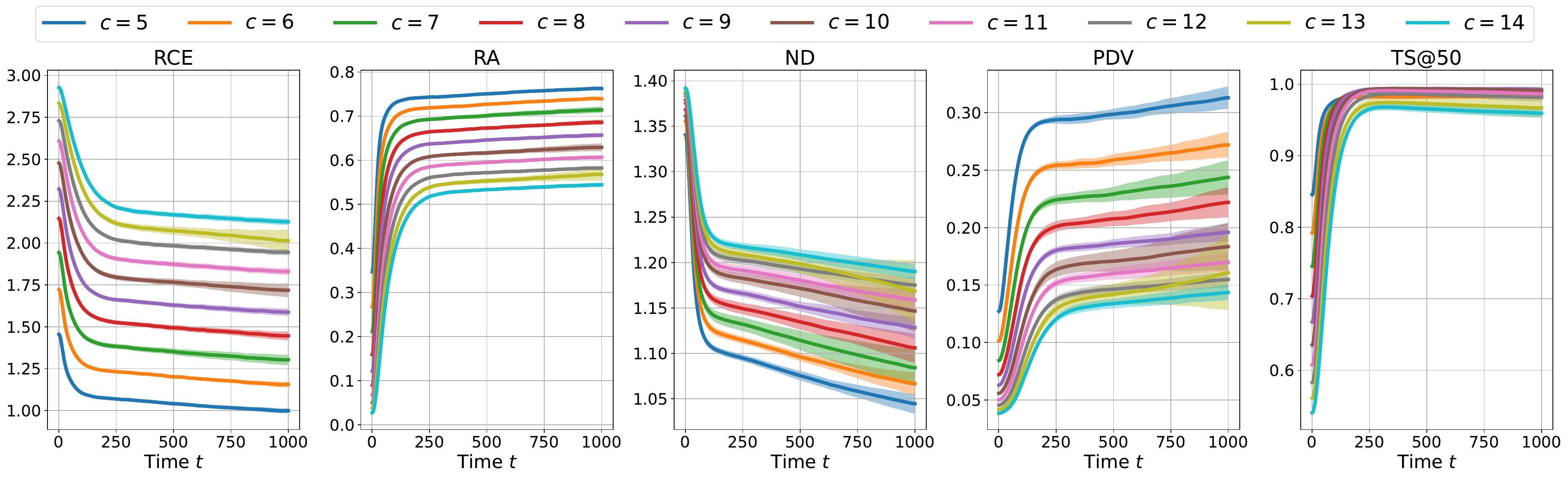}
  \caption{Metric trends over time under varying $c$ on Synthetic dataset.}
  \label{fig:synthetic_c}
\end{figure*}
\paragraph{\textbf{Sensitivity Effect $\boldsymbol{c}$.}}
As the number of item categories $c$ increases in Figure \ref{fig:synthetic_c}: 1) Both RCE and ND increase. 2) PDV and TS decrease. This indicates that increasing the number of item categories can help mitigate echo chamber, social homogenization, global polarization, and localized homogenization. Increasing the number of item categories directly enhances the category diversity in recommendation lists and expands the range of user interests, thereby alleviating homogenization.

In summary, our results show that increasing items and categories helps reduce echo chambers and homogenization by diversifying recommendations and user interests, and higher social network density reduces social homogenization but can increase global polarization. These insights emphasize how dataset scale  influence echo chambers and user homogenization.

\begin{figure}[!h]
  \centering
  \includegraphics[width=1\linewidth]{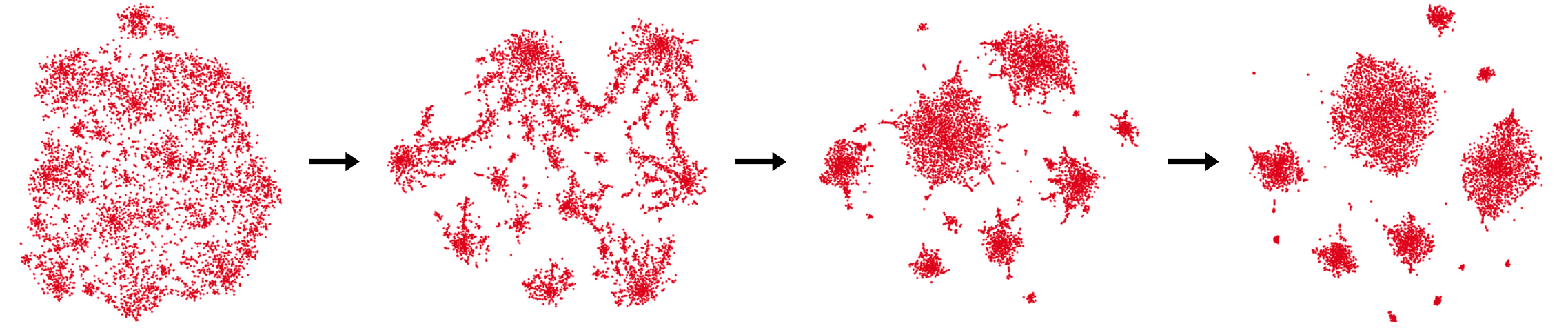}
  \caption{Temporal evolution of user representations visualization.}
  \label{fig:ciao_tsne}
  \vspace{-15pt}
\end{figure}

\subsubsection{Visualization of User Representations.}
We visualize the temporal evolution of user representations using t-SNE \cite{van2008visualizing} on the Ciao dataset, under the parameters $\alpha=5$, $\beta=5$, $\gamma=0.5$, and $\epsilon=0$, as shown in Figure \ref{fig:ciao_tsne}. Although we initialize user interest representations from their interaction history, a clear pattern of local clustering still emerges over time, which aligns with the trends observed in our proposed metrics and indicates the presence of both global polarization and local homogenization.

\subsubsection{Additional Experiments.}
In Appendix \ref{apd:para-anal}, we analyze the impact of parameters on the Epinions and synthetic datasets.

\subsection{Mitigation and Results}
We compute the dispersion of a user's interest representation based on the $\ell_2$-normalized user representation $\mathbf{u}_i(t)$, denoted as $\mathbf{u}_i^{dis}(t)=\|\mathbf{u}_i(t)-\mu_i(t)\|_2^2$, where $\mu_i(t)=\frac{\sum_{k=1}^c u_i^{(k)}(t)}{c}$. A smaller $\mathbf{u}_i^{dis}(t)$ indicates a broader and more diverse range of interests, whereas a larger $\mathbf{u}_i^{dis}(t)$
reflects more extreme and concentrated preferences. 

\paragraph{\textbf{(1)\ User Adaptive $\boldsymbol{\alpha}$ (UA$\boldsymbol{\alpha}$).}}
Under a global stochasticity level $\alpha_0$ in the recommendation process, we adaptively adjust the $\alpha_i(t)$ for each user $i$ at time $t$. Specifically, for users with narrower interest representations, we slightly reduce $\alpha_i(t)$ to encourage more diverse recommendations; conversely, for users with broader interests, we increase $\alpha_i(t)$ to maintain recommendation relevance.
\begin{equation}
\alpha_i(t)=\frac{\phi_i(t;\sigma)}{\sum_{k=1}^n\phi_k(t;\sigma)}\alpha_0
\end{equation}
where \(\phi_i(t;\sigma)={(\frac{1}{\mathbf{u}_i^{dis}(t)})}^\sigma\), and $\sigma$ serves as a temperature parameter.

\paragraph{\textbf{(2)\ Feedback Update Adjustment (FUA)}}
As suggested in \cite{piao2023human}, placing more emphasis on negative feedback can help alleviate the decline in recommendation diversity. Therefore, we modify the weights of positive and negative feedback in Equation \ref{eq:user-update}:
\begin{equation}
w_i^t(r_i^j) = 
\begin{cases}
1-\rho, & \text{with probability } p_{pos}^t(\mathbf{u}_i, \mathbf{v}_{r_i^j}; \beta, \epsilon) \\
-1-\rho, & \text{with probability } p_{neg}^t(\mathbf{u}_i, \mathbf{v}_{r_i^j}; \beta, \epsilon)
\end{cases}
\end{equation}
where the parameter $\rho > 0$ increases the impact of negative feedback while reducing the influence of positive feedback.

\paragraph{\textbf{(3)\ Diversity Post-Processing (DPP)}}
We enhance the diversity of the recommendation list by adopting a post-processing strategy similar to \cite{carbonell1998use}, described as follows. First, We sample 1000 items without replacement from the probability distribution based on Equation (3), consistent with the recommendation procedure described above, to construct the candidate set $\mathcal{D}_i(t)$ for user $i$ at time $t$. Next, we select the most relevant item $l_1^i$ from $\mathcal{D}_i(t)$ to initialize the re-ranked list $\mathcal{L}_i(t)$. Then we select the next item $l_j^i$ using a diversity-penalized strategy: 

\begin{table*}[!t]
  \caption{Improvements of the original metrics by the four mitigation strategies on Ciao and Epinions datasets.}
  \centering
  \renewcommand{\arraystretch}{1.05}
  \setlength{\tabcolsep}{4pt}
  \fontsize{9.5}{10.5}\selectfont

  \resizebox{\textwidth}{!}{
  \begin{tabular}{|>{\centering\arraybackslash}p{2.5cm}|ccccc|ccccc|}
    \toprule
    \multirow{2}{*}{\textbf{Methods}} & \multicolumn{5}{c|}{\textbf{Ciao}} & \multicolumn{5}{c|}{\textbf{Epinions}} \\
    & RCE $\uparrow$ & RA $\uparrow$ & ND $\uparrow$ & PDV $\downarrow$ & TS@300 $\downarrow$ & RCE $\uparrow$ & RA $\uparrow$ & ND $\uparrow$ & PDV $\downarrow$ & TS@900 $\downarrow$ \\
    \midrule
    Original & 1.20$_{\pm 0.002}$ & 0.74$_{\pm 0.001}$ & 0.88$_{\pm 0.005}$ & 0.31$_{\pm 0.002}$ & 0.96$_{\pm 0.000}$ & 1.35$_{\pm 0.001}$ & 0.71$_{\pm 0.000}$ & 0.74$_{\pm 0.002}$ & 0.46$_{\pm 0.001}$ & 0.95$_{\pm 0.000}$ \\
    \midrule
    UA$\alpha$ & 1.28$_{\pm 0.000}$ & 0.72$_{\pm 0.000}$ & 0.95$_{\pm 0.002}$ & 0.28$_{\pm 0.002}$ & 0.96$_{\pm 0.001}$ & 1.44$_{\pm 0.004}$ & 0.68$_{\pm 0.001}$ & 0.75$_{\pm 0.010}$ & 0.46$_{\pm 0.002}$ & 0.95$_{\pm 0.000}$ \\
    Improv.\% ($p$-val) & 6.55\% (0.00) & -2.38\% (0.00) & 7.95\% (0.00) & 9.93\% (0.01) & 0.39\% (0.09) & 6.66\% (0.00) & -3.71\% (0.00) & 2.40\% (0.10) & 2.05\% (0.03) & 0.34\% (0.00) \\
    \midrule
    FUA & 1.27$_{\pm 0.009}$ & 0.73$_{\pm 0.002}$ & 0.92$_{\pm 0.007}$ & 0.29$_{\pm 0.005}$ & 0.96$_{\pm 0.000}$ & 1.43$_{\pm 0.004}$ & 0.69$_{\pm 0.001}$ & 0.83$_{\pm 0.007}$ & 0.46$_{\pm 0.001}$ & 0.95$_{\pm 0.000}$ \\
    Improv.\% ($p$-val) & 5.73\% (0.00) & -2.10\% (0.00) & 4.86\% (0.02) & 5.42\% (0.01) & 0.37\% (0.00) & 6.35\% (0.00) & -2.73\% (0.00) & 13.04\% (0.00) & 1.34\% (0.00) & 0.60\% (0.00) \\
    \midrule
    DPP & 1.38$_{\pm 0.001}$ & 0.74$_{\pm 0.000}$ & 0.89$_{\pm 0.006}$ & 0.30$_{\pm 0.003}$ & 0.96$_{\pm 0.000}$ & 1.45$_{\pm 0.003}$ & 0.73$_{\pm 0.001}$ & 0.71$_{\pm 0.009}$ & 0.47$_{\pm 0.001}$ & 0.95$_{\pm 0.000}$ \\
    Improv.\% ($p$-val) & 15.46\% (0.00) & -0.63\% (0.01) & 0.96\% (0.02) & 1.38\% (0.07) & 0.00\% (0.94) & 7.36\% (0.00) & 3.34\% (0.00) & -3.88\% (0.05) & -0.40\% (0.19) & -0.03\% (0.56) \\
    \midrule
    SAR & 1.26$_{\pm 0.009}$ & 0.68$_{\pm 0.003}$ & 0.94$_{\pm 0.008}$ & 0.28$_{\pm 0.004}$ & 0.95$_{\pm 0.001}$ & 1.51$_{\pm 0.007}$ & 0.67$_{\pm 0.002}$ & 0.96$_{\pm 0.034}$ & 0.41$_{\pm 0.003}$ & 0.94$_{\pm 0.001}$ \\
    Improv.\% ($p$-val) & 5.47\% (0.01) & -7.88\% (0.00) & 6.53\% (0.02) & 9.95\% (0.00) & 1.03\% (0.00) & 11.79\% (0.00) & -4.85\% (0.00) & 31.02\% (0.01) & 12.74\% (0.00) & 1.04\% (0.01) \\
    \bottomrule
  \end{tabular}}
\end{table*}

\begin{equation}
l_j^i = \operatorname*{argmax}_{l \in \mathcal{D}_i(t) \setminus \{l_1^i,l_2^i,\ldots,l_{j-1}^i\}} \left\{ (1-\theta) \mathbf{u}_i(t)^T\mathbf{v}_l  - \theta \mathbf{v}_l^T \frac{\sum_{k=1}^{j-1}\mathbf{v}_{l_k^i}}{\|\sum_{k=1}^{j-1}\mathbf{v}_{l_k^i}\|_2} \right\}
\end{equation}until the re-ranked list $\mathcal{L}_i(t)$ contains 
$h=20$ items, where $\theta$ controls the level of diversity.
\paragraph{\textbf{(4)\ Social Aggregation Reweighting (SAR)}}
To reduce the impact of users with extreme interests and alleviate global polarization and local homogenization, we apply a reweighting strategy when aggregating neighbor preferences during the recommendation process in Equation \ref{eq:social-similar-match}:
\begin{equation}
\mathbf{s}_i^{\gamma}(t) = \gamma \mathbf{u}_i(t) + (1 - \gamma) \frac{ \sum_{j \in \mathcal{N}_i } \exp(-\omega \mathbf{u}_j^{dis}(t))  \mathbf{u}_j(t) }{ \sum_{j \in \mathcal{N}_i } \exp(-\omega \mathbf{u}_j^{dis}(t))|\mathcal{N}_i| }
\end{equation}where $\omega$ serves as a temperature parameter.

We compare the performance improvements of all mitigation strategies over the original metrics on the Ciao and Epinions datasets. Each reported metric represents the average over 1,000 time steps $t$, with three runs under different seeds to compute standard deviations and $p$-values. As shown in Table \ref{main experiment}, the experimental configuration for Ciao is: $\sigma=10$, $\rho=0.02$, $\theta=0.501$, $\omega=1000$, while for Epinions it is: $\sigma=10$, $\rho=0.02$, $\theta=0.501$, $\omega=150$. Since all strategies involve tunable intensity parameters, we further analyze their influence on the results in Appendix \ref{apd:mitigate}, based on the Ciao dataset.

The results demonstrate that all four mitigation strategies can effectively improve recommendation diversity and alleviate user homogenization, albeit with a slight trade-off in accuracy. Among them, DPP achieves the best performance in enhancing diversity, confirming the effectiveness of post-processing techniques in recommender systems \cite{ziegler2005improving}. SAR performs best in mitigating both polarization and homogenization, highlighting the potential of improving social recommendation by refining neighbor aggregation \cite{sharma2024survey}. UA$\alpha$ suggests that recommendation strategies should adaptively adjust to individual user behaviors for controllable personalization \cite{wang2022user}. Finally, FUA supports the conclusion in \cite{piao2023human} that better leveraging negative feedback can lead to unexpectedly outcomes.

\section{Discussion}
\label{sec:conclusion}
In this work, we present a novel confirmation-aware social dynamic model that systematically investigates the emergence and intensification of echo chambers and user homogenization in recommender systems. By integrating user psychological mechanisms (e.g., confirmation bias, leniency bias) and social network dynamics into a closed-loop interaction framework, our model provides theoretical and empirical insights into how algorithmic recommendations and human behavior co-evolve toward systemic distortions. Theoretically, we establish that reduced recommendation diversity (echo chambers) and homogenized user representations are inevitable under mild assumptions about user psychology and social connectivity. Empirically, experiments on real-world and synthetic datasets demonstrate that three key dimensions — system-level (recommendation stochasticity and sociality), user-level (user psychology), and platform-level (dataset scale) — significantly modulate these phenomena. Specifically, increasing recommendation accuracy and intensifying confirmation and leniency biases tend to exacerbate echo chambers and homogenization by reinforcing preference-aligned feedback, while introducing more social information into recommendation or providing users with a greater number of items and item categories helps alleviate these effects. Furthermore, our proposed mitigation strategies offer practical pathways for human-centered recommendation design. These findings advance the understanding of algorithmic feedback loops and provide actionable guidelines for balancing recommendation relevance with societal well-being.

\paragraph{Limitations.}
\label{apd:limitations}

While our dynamic model captures core interaction mechanisms between users and recommendation systems, it simplifies the complexities arising from unobservable factors, unpredictable behaviors, and the real-world update mechanisms of recommendation algorithms. Moreover, certain assumptions are made in the derivation of user homogenization and entropy decline. This study prioritizes establishing a theoretical foundation to demonstrate the inherent tendencies toward echo chambers and user homogenization, while systematically identifying their key drivers through controlled experiments. The experiments provide partial evidence supporting the validity of our theoretical framework, future work should focus on three directions: (1) enhancing simulation realism using advanced techniques like large language models (LLMs) to better mimic user decision-making, (2) expanding impact assessments to evaluate broader societal consequences, (3) verifying the validity of our conclusions and the effectiveness of mitigation strategies under real-world recommender systems.

%% file: appendix.tex
\section{Related Work}
\label{apd:related work}

\begin{table}[h]
\centering
\caption{Comparison between our work and some previous works.}
\label{apd:related work table}
\scalebox{0.78}{
\begin{tabular}{|l|c|c|c|c|}
\hline
\textbf{Works} & \textbf{\begin{tabular}[c]{@{}l@{}}With Recomme\\-nder Systems?\end{tabular}} & \textbf{\begin{tabular}[c]{@{}l@{}}With User\\ Psychology?\end{tabular}} & \textbf{\begin{tabular}[c]{@{}l@{}}With Social\\ Network?\end{tabular}}   & \textbf{\begin{tabular}[c]{@{}l@{}}Explores Under\\-lying Causes?\end{tabular}}                                                                               \\ \hline

\cite{chaney2018algorithmic}   & \ding{51}   & \ding{55}   & \ding{55}  & \ding{55}  \\ \hline
\cite{mansoury2020feedback}   & \ding{51}   & \ding{55}   & \ding{55}  & \ding{55}  \\ \hline
\cite{sun2019debiasing}   & \ding{51}   & \ding{55}   & \ding{55}  & \ding{55}  \\ \hline
\cite{jiang2019degenerate}   & \ding{51}   & \ding{55}   & \ding{55}  & \ding{51}  \\ \hline
\cite{kalimeris2021preference}   & \ding{51}   & \ding{55}   & \ding{55}  & \ding{51}  \\ \hline
\cite{piao2023human}   & \ding{51}   & \ding{55}   & \ding{55}  & \ding{51}  \\ \hline
\cite{chen2024coevolution}   & \ding{51}   & \ding{55}   & \ding{51}  & \ding{55}  \\ \hline
\cite{rossi2021closed}   & \ding{51}   & \ding{55}   & \ding{55}  & \ding{55}  \\ \hline
\cite{tornberg2022digital}   & \ding{55}   & \ding{55}   & \ding{51}  & \ding{55}  \\ \hline
\cite{banisch2019opinion}   & \ding{55}   & \ding{55}   & \ding{51}  & \ding{55}  \\ \hline
\cite{perra2019modelling}   & \ding{55}   & \ding{55}   & \ding{51}  & \ding{51}  \\ \hline
\cite{chitra2020analyzing}   & \ding{55}   & \ding{55}   & \ding{51}  & \ding{55}  \\ \hline
Ours & \textcolor{green}{\ding{51}} & \textcolor{green}{\ding{51}} & \textcolor{green}{\ding{51}} & \textcolor{green}{\ding{51}} \\ \hline
\end{tabular}}

\vspace{0.2em}

\end{table}

In recent years, echo chambers \cite{liu2021interaction, mckay2022turn} and user homogenization \cite{chaney2018algorithmic, mansoury2020feedback} in recommender systems have attracted widespread attention. Some studies have statistically confirmed the existence of these effects but fall short of investigating their underlying causes and the dynamic interaction mechanisms that drive them \cite{nguyen2014exploring, ge2020understanding, levy2021social}. However, modern recommender systems \cite{he2020lightgcn, zhang2019deep} are typically built upon deep learning architectures, whose black-box nature and heavy reliance on numerous parameters make it difficult to explain the essential reasons behind these phenomena \cite{rudin2019stop, zhang2020explainable}.

\paragraph{Echo Chambers and Homogenization through User–RecSys Feedback Loops.}
Several works have proposed dynamic models that simulate iterative interactions between users and recommender systems \cite{chaney2018algorithmic, mansoury2020feedback, sun2019debiasing, jiang2019degenerate, kalimeris2021preference, piao2023human}, aiming to explore the mechanisms behind the emergence of echo chambers. \cite{chaney2018algorithmic} found that the recommendation feedback loop leads to homogenization of user behavior and causes users to experience utility loss, offering insights for the design of real-world recommender systems. \cite{mansoury2020feedback} found that the iterative interaction between recommendation algorithms and users amplifies bias, resulting in decreased diversity and user homogenization. \cite{sun2019debiasing} found that the feedback loop in the iterative interaction between matrix factorization algorithms and users amplifies popularity bias, and proposed improved debiasing algorithms to mitigate this effect. \cite{jiang2019degenerate} modeled the evolution of user interests using a dynamical system framework, and discussed how three key design factors of the model influence the emergence of echo chamber effects. \cite{kalimeris2021preference} modeled the dynamic interactions between users and recommender systems, validated the extremization of user preference representations that leads to echo chambers, and proposed corresponding mitigation strategies. \cite{piao2023human} theoretically and empirically validated the information cocoon phenomenon using an adaptive information dynamics model, revealing that positive feedback drives entropy reduction, while negative feedback helps counteract the effect. Nevertheless, these models often overlook user-side psychological biases and the pervasive influence of social networks in real-world settings. 

\paragraph{Social Polarization through Opinion Dynamics.}
Other studies have drawn from opinion dynamics to simulate how users update their views within social networks \cite{tornberg2022digital, banisch2019opinion, perra2019modelling, chitra2020analyzing}, thereby analyzing the evolution of user opinions. \cite{tornberg2022digital} validated the echo chamber phenomenon on digital media using opinion dynamics models, revealing its role in driving affective polarization and societal division. \cite{banisch2019opinion} explored a novel mechanism of polarization in opinion dynamics based on social feedback, showing that in sufficiently modular networks, distinct groups of agents can develop strong, opposing convictions. \cite{perra2019modelling} developed an opinion dynamics model and finds that algorithmic filtering mechanisms can intensify echo chambers and polarization in socially structured networks, while heterogeneity in connectivity helps mitigate these effects. \cite{chitra2020analyzing} extended the Friedkin-Johnsen opinion dynamics model and found that reducing disagreement among social neighbors significantly amplifies user opinion polarization. However, these approaches usually detach from the recommender system context, focusing solely on interpersonal influence. 

More importantly, most of the above works primarily provide empirical descriptions of these phenomena, with limited exploration of the underlying driving factors behind their emergence. In contrast to these works, our study incorporates social networks into the user–recommender interaction loop, extending the analysis to a more realistic social recommendation scenario. Moreover, we model user psychological mechanisms—such as confirmation bias and leniency bias—to better capture real-world feedback behaviors, and further identify and analyze the root causes that drive the emergence of these phenomena. Table \ref{apd:related work table} highlights the differences between several prior dynamic modeling approaches and our work.

\section{Experimental Settings}
\label{apd:implementation}

We vary the parameters $\alpha$, $\beta$, $\gamma$ and $\epsilon$ individually to observe the temporal dynamics of the above metrics, while keeping the other three fixed. The default values for these parameters are set to $\alpha = 5$, $\beta = 5$, $\gamma = 0.5$, and $\epsilon = 0$. The recommendation list length is fixed at $|\mathcal{R}_i^t| = h = 20$, and the values of $k$ for the $TS@k$ metric are set to 300, 900, and 50 for the Ciao, Epinions, and synthetic datasets, respectively, in accordance with the dataset sizes, as larger user numbers require larger $k$ to capture $TS$ metric value differences effectively. For real-world datasets, each experiment is repeated with three different random seeds, while for the synthetic dataset, we conduct 10 repeated runs. We plot the mean curves along with the 95\% confidence intervals.

\section{Additional Experiments on Parameter Analysis}
\label{apd:para-anal}
We perform the same analysis of the four factors on the Epinions and synthetic datasets, and the results show a high degree of consistency with those on the Ciao dataset.
\subsection{Analysis of $\alpha$, $\beta$, $\gamma$ 
 and $\epsilon$ on Epinions Dataset}
\begin{figure}[H]
  \centering
  \includegraphics[width=1\linewidth]{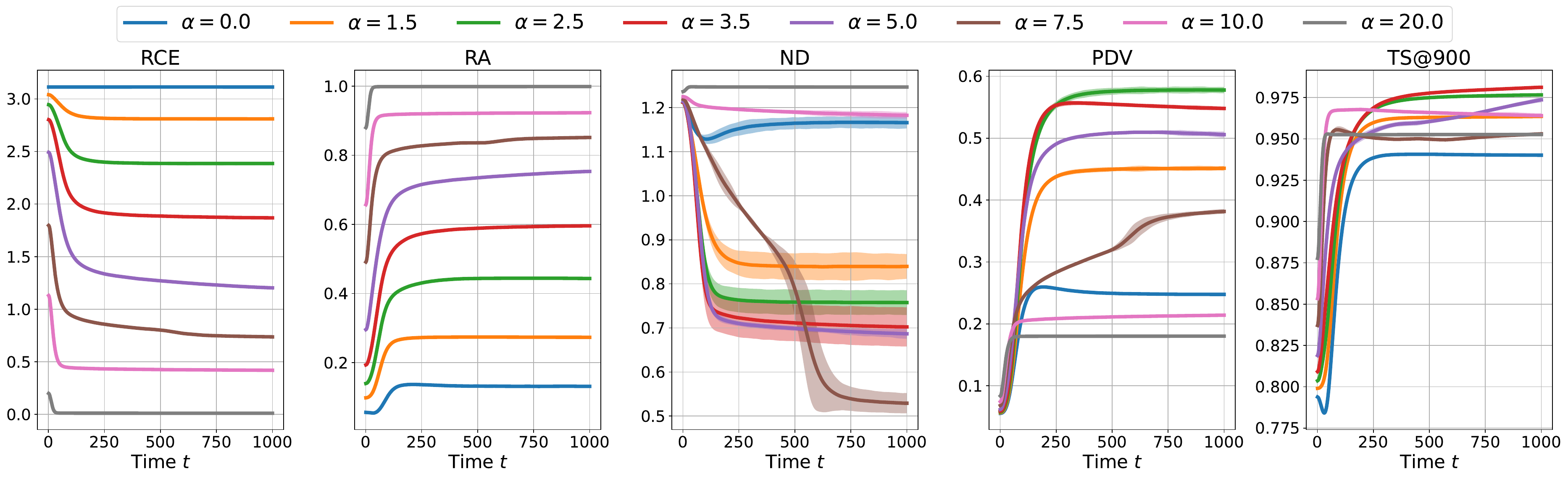}
  \caption{Metric trends over time under varying $\alpha$ on Epinions dataset.}
  \label{fig:epinions_alpha}
  \vspace{-15pt}
\end{figure}

\begin{figure}[H]
  \centering
  \includegraphics[width=1\linewidth]{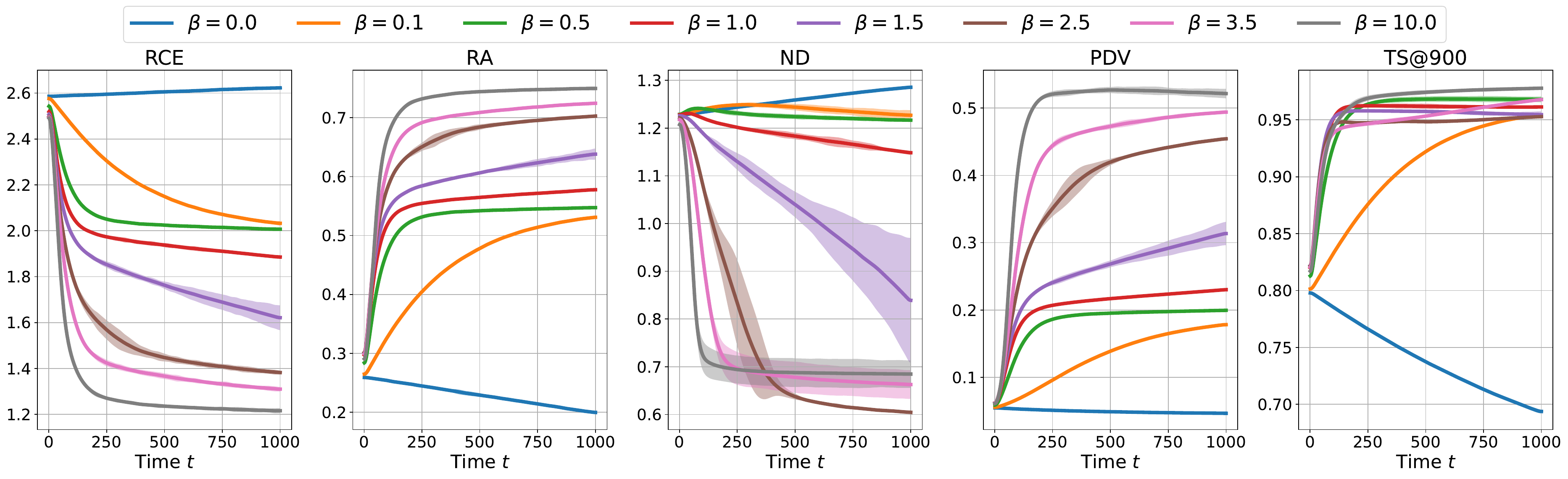}
  \caption{Metric trends over time under varying $\beta$ on Epinions dataset.}
  \label{fig:epinions_beta}
  \vspace{-15pt}
\end{figure}

\begin{figure}[H]
  \centering
  \includegraphics[width=1\linewidth]{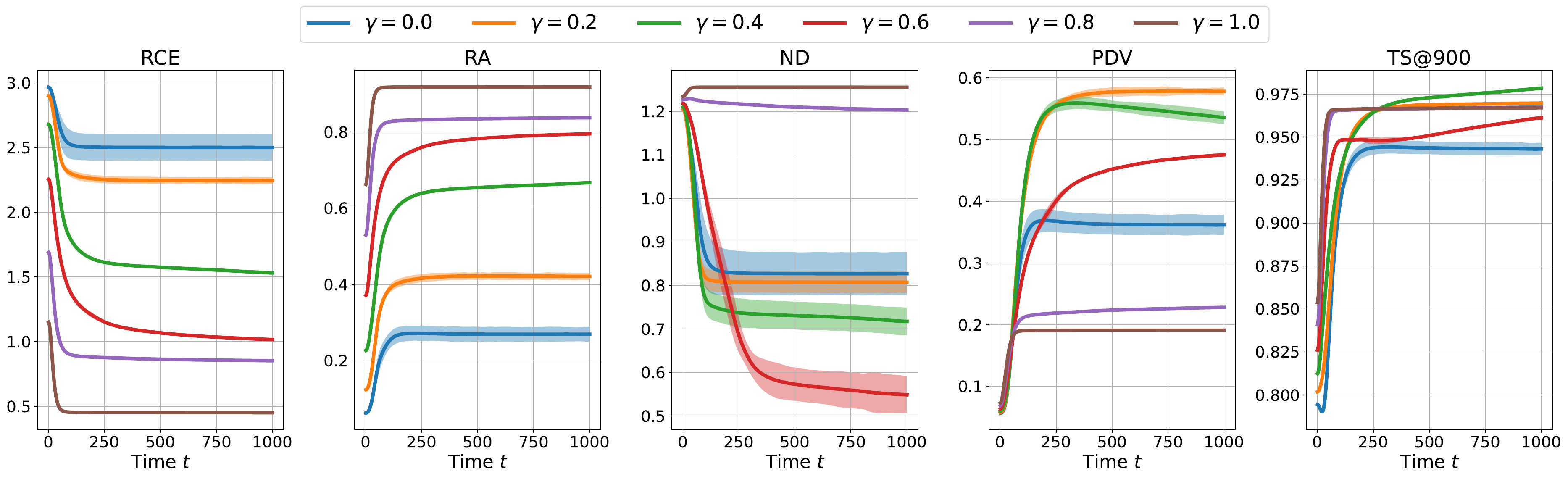}
  \caption{Metric trends over time under varying $\gamma$ on Epinions dataset.}
  \label{fig:epinions_gamma}
  \vspace{-15pt}
\end{figure}

\begin{figure}[H]
  \centering
  \includegraphics[width=1\linewidth]{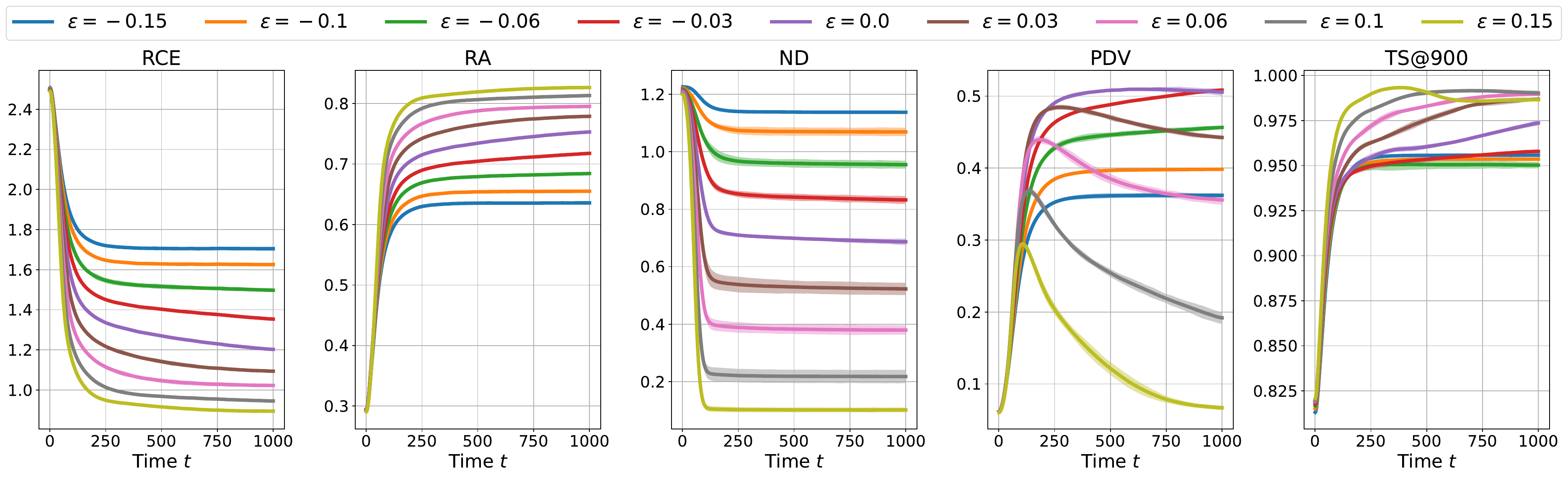}
  \caption{Metric trends over time under varying $\epsilon$ on Epinions dataset.}
  \label{fig:epinions_epsilon}
  \vspace{-15pt}
\end{figure}
\subsection{Analysis of $\alpha$, $\beta$, $\gamma$ 
 and $\epsilon$ on Synthetic Dataset}

\begin{figure}[H]
  \centering
  \includegraphics[width=1\linewidth]{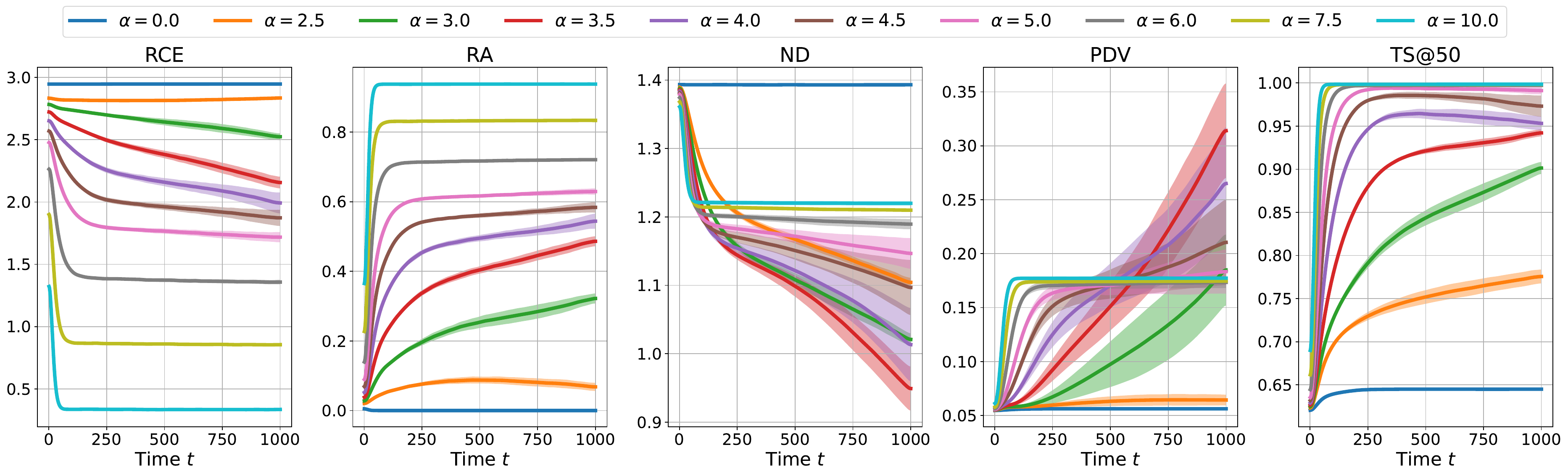}
  \caption{Metric trends over time under varying $\alpha$ on Synthetic dataset.}
  \label{fig:synthetic_alpha}
  \vspace{-15pt}
\end{figure}

\begin{figure}[H]
  \centering
  \includegraphics[width=1\linewidth]{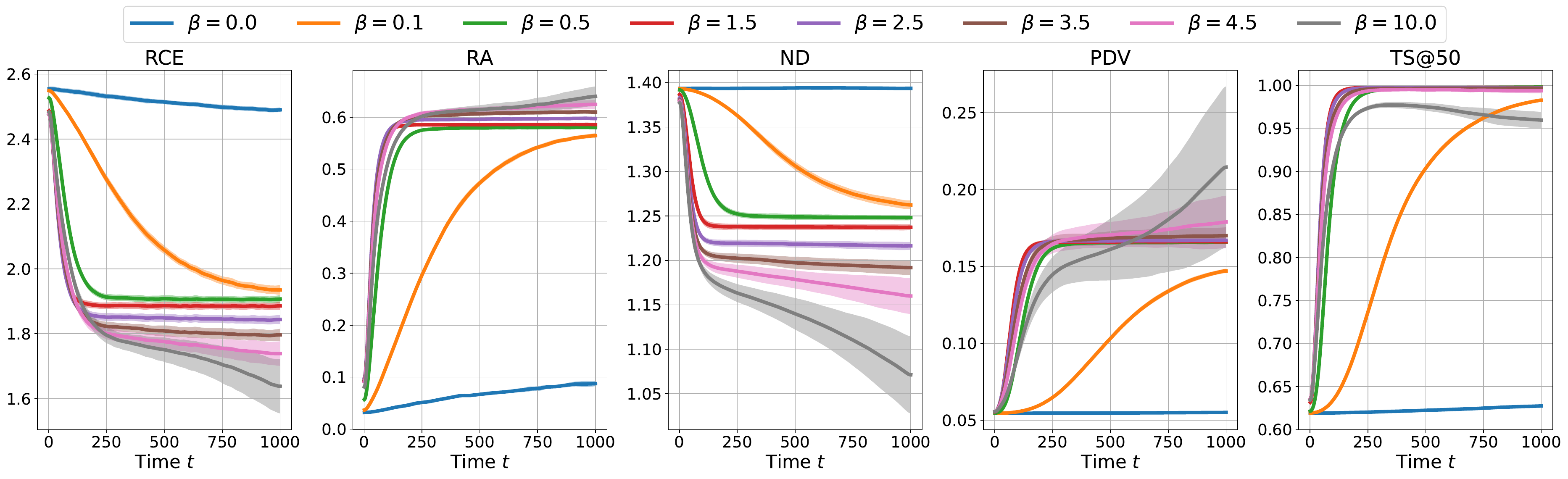}
  \caption{Metric trends over time under varying $\beta$ on Synthetic dataset.}
  \label{fig:synthetic_beta}
  \vspace{-15pt}
\end{figure}

\begin{figure}[H]
  \centering
  \includegraphics[width=1\linewidth]{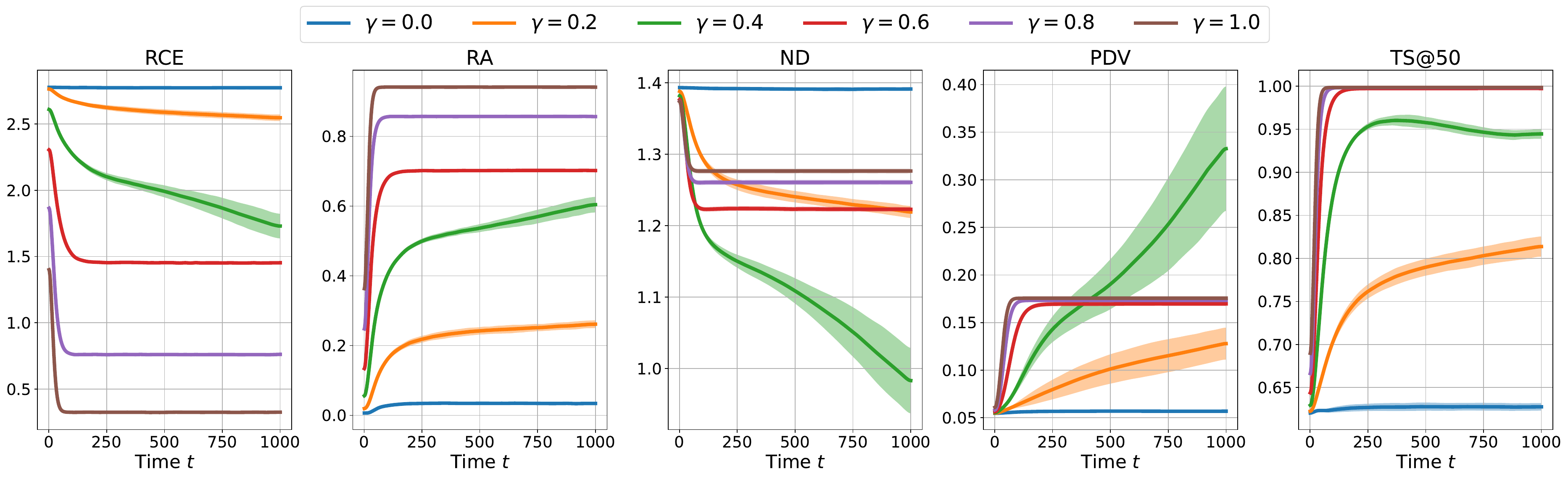}
  \caption{Metric trends over time under varying $\gamma$ on Synthetic dataset.}
  \label{fig:synthetic_gamma}
  \vspace{-15pt}
\end{figure}

\begin{figure}[H]
  \centering
  \includegraphics[width=1\linewidth]{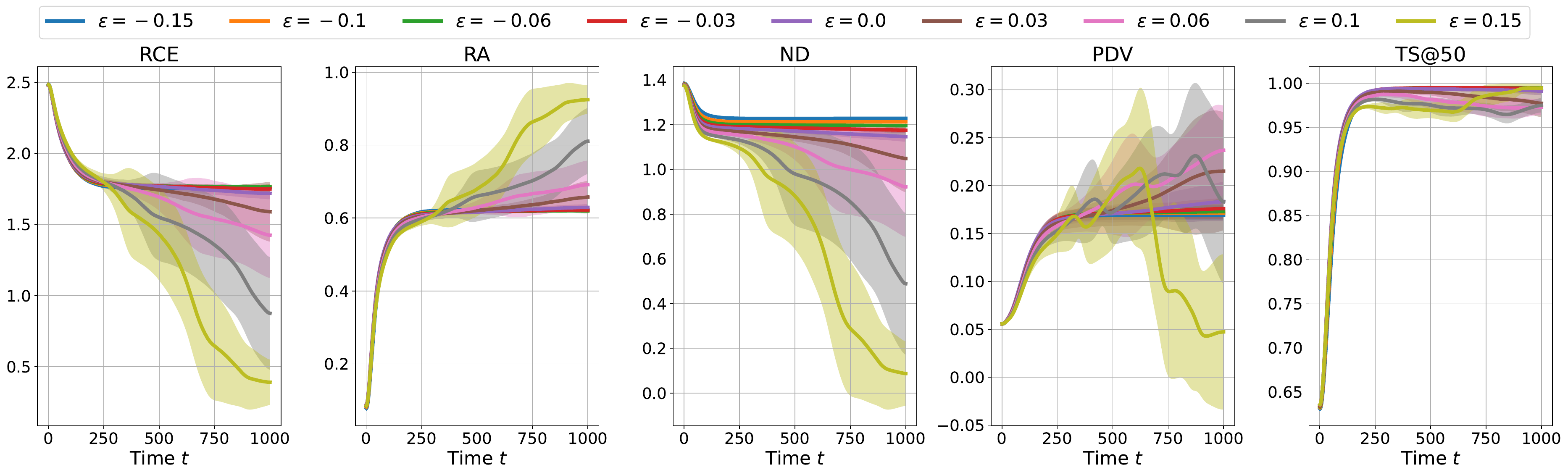}
  \caption{Metric trends over time under varying $\epsilon$ on Synthetic dataset.}
  \label{fig:synthetic_epsilon}
  \vspace{-15pt}
\end{figure}

\begin{table*}[!h]
    \newcommand{\pv}[1]{{\scriptsize \text{#1}}} 
    \centering
    \caption{Parameter sensitivity analysis of four strategies on Ciao dataset.}
    \label{main experiment}
    \renewcommand{\arraystretch}{1} 
    \vspace{-5pt}
    \resizebox{0.6\linewidth}{!}{
    \begin{tabular}{|cc|ccccc|}
    \toprule
    \multicolumn{2}{|c|}{\text{Ciao}} & RCE$\uparrow$ & RA$\uparrow$ & ND$\uparrow$ & PDV$\downarrow$ & TS@300$\downarrow$ \\
    \midrule
    \multirow{4}{*}{UA$\alpha$} & 
    $\alpha=1$ & $\text{1.21}_{\pm \text{0.004}}$ & $\text{0.74}_{\pm \text{0.001}}$ & $\text{0.90}_{\pm \text{0.005}}$ & $\text{0.30}_{\pm \text{0.002}}$ & $\text{0.96}_{\pm \text{0.001}}$  \\
    & $\alpha=5$ & $\text{1.24}_{\pm \text{0.003}}$ & $\text{0.73}_{\pm \text{0.001}}$ & $\text{0.94}_{\pm \text{0.001}}$ & $\text{0.29}_{\pm \text{0.002}}$ & $\text{0.96}_{\pm \text{0.001}}$  \\
    & $\alpha=10$ & $\text{1.28}_{\pm \text{0.000}}$ & $\text{0.72}_{\pm \text{0.000}}$ & $\text{0.95}_{\pm \text{0.002}}$ & $\text{0.28}_{\pm \text{0.002}}$ & $\text{0.96}_{\pm \text{0.001}}$  \\
    & $\alpha=20$ & $\text{1.32}_{\pm \text{0.004}}$ & $\text{0.71}_{\pm \text{0.001}}$ & $\text{0.98}_{\pm \text{0.001}}$ & $\text{0.27}_{\pm \text{0.000}}$ & $\text{0.95}_{\pm \text{0.000}}$  \\

    \midrule
    \multirow{4}{*}{FUA} & 
    $\rho=-0.01$ & $\text{1.16}_{\pm \text{0.004}}$ & $\text{0.75}_{\pm \text{0.001}}$ & $\text{0.86}_{\pm \text{0.004}}$ & $\text{0.32}_{\pm \text{0.003}}$ & $\text{0.96}_{\pm \text{0.001}}$  \\
    & $\rho=0.02$ & $\text{1.27}_{\pm \text{0.009}}$ & $\text{0.73}_{\pm \text{0.002}}$ & $\text{0.92}_{\pm \text{0.007}}$ & $\text{0.29}_{\pm \text{0.005}}$ & $\text{0.96}_{\pm \text{0.000}}$  \\
    & $\rho=0.05$ & $\text{1.36}_{\pm \text{0.005}}$ & $\text{0.70}_{\pm \text{0.001}}$ & $\text{0.96}_{\pm \text{0.003}}$ & $\text{0.27}_{\pm \text{0.002}}$ & $\text{0.95}_{\pm \text{0.000}}$  \\
    & $\rho=0.08$ & $\text{1.44}_{\pm \text{0.004}}$ & $\text{0.68}_{\pm \text{0.001}}$ & $\text{1.00}_{\pm \text{0.001}}$ & $\text{0.26}_{\pm \text{0.002}}$ & $\text{0.95}_{\pm \text{0.001}}$  \\

    \midrule
    \multirow{4}{*}{DPP} & 
    $\theta=0.500$ & $\text{1.13}_{\pm \text{0.001}}$ & $\text{0.78}_{\pm \text{0.000}}$ & $\text{0.89}_{\pm \text{0.006}}$ & $\text{0.30}_{\pm \text{0.003}}$ & $\text{0.96}_{\pm \text{0.000}}$  \\
    & $\theta=0.501$ & $\text{1.38}_{\pm \text{0.001}}$ & $\text{0.74}_{\pm \text{0.000}}$ & $\text{0.97}_{\pm \text{0.004}}$ & $\text{0.27}_{\pm \text{0.002}}$ & $\text{0.95}_{\pm \text{0.000}}$  \\
    & $\theta=0.502$ & $\text{1.61}_{\pm \text{0.001}}$ & $\text{0.70}_{\pm \text{0.000}}$ & $\text{1.04}_{\pm \text{0.003}}$ & $\text{0.24}_{\pm \text{0.003}}$ & $\text{0.94}_{\pm \text{0.001}}$  \\
    & $\theta=0.503$ & $\text{1.80}_{\pm \text{0.001}}$ & $\text{0.66}_{\pm \text{0.000}}$ & $\text{1.09}_{\pm \text{0.002}}$ & $\text{0.22}_{\pm \text{0.001}}$ & $\text{0.94}_{\pm \text{0.001}}$  \\

    \midrule
    \multirow{4}{*}{SAR} & 
    $\omega=100$ & $\text{1.22}_{\pm \text{0.005}}$ & $\text{0.72}_{\pm \text{0.001}}$ & $\text{0.85}_{\pm \text{0.016}}$ & $\text{0.31}_{\pm \text{0.004}}$ & $\text{0.96}_{\pm \text{0.002}}$  \\
    & $\omega=1000$ & $\text{1.26}_{\pm \text{0.009}}$ & $\text{0.68}_{\pm \text{0.003}}$ & $\text{0.94}_{\pm \text{0.008}}$ & $\text{0.28}_{\pm \text{0.004}}$ & $\text{0.95}_{\pm \text{0.001}}$  \\
    & $\omega=2000$ & $\text{1.24}_{\pm \text{0.003}}$ & $\text{0.69}_{\pm \text{0.001}}$ & $\text{0.90}_{\pm \text{0.019}}$ & $\text{0.29}_{\pm \text{0.005}}$ & $\text{0.95}_{\pm \text{0.000}}$  \\
    & $\omega=5000$ & $\text{1.25}_{\pm \text{0.004}}$ & $\text{0.68}_{\pm \text{0.001}}$ & $\text{0.92}_{\pm \text{0.013}}$ & $\text{0.28}_{\pm \text{0.004}}$ & $\text{0.95}_{\pm \text{0.001}}$  \\
    \bottomrule
    \end{tabular}}
\end{table*}
\section{Additional Experiments with Mitigation Strategies on Ciao Dataset}
\label{apd:mitigate}
Through further parameter tuning of the four mitigation strategies on the Ciao dataset in Table \ref{main experiment}, we observe that sacrificing a certain degree of accuracy can lead to more effective alleviation of echo chambers and user homogenization.

\section{Proof of Proposition \ref{thm:mat-format}}
\label{apd:thm1}
\begin{proof}
\ 
\begin{itemize}[noitemsep,topsep=0pt,parsep=2pt,partopsep=0pt, leftmargin=1.5em]
\item First we define the indicator function as follows:

\begin{equation}
\mathbf{1}^t_i(j) = 
\begin{cases} 
1, & \parbox[t]{.7\linewidth}{
\raggedright if item $j$ appears in $h$ samples drawn without replacement from $\mathbf{p}_{\mathbf{s}_i}^{t}(\alpha)$
} \\
0, & \text{otherwise}
\end{cases}
\end{equation}

Based on Lemma \ref{lm:1}, we have 
\(
\mathbb{E} \left[  \mathbf{1}^t_i(j)\right] \approx hp^{t}(\mathbf{s}_i^{\gamma}, \mathbf{v}_j; \alpha)
\), 
then Equation \ref{eq:user-update} admits the following expectation representation:

\begin{align}\label{eq:pg}
&\mathbf{u}_i(t+1) = \mathbf{u}_i(t) + \frac{\eta}{h} \mathbb{E}\left[ \sum_{j=1}^m  \mathbf{1}^t_i(j)w_i^t(j)\mathbf{v}_{j}\right] \nonumber\\
&=\mathbf{u}_i(t) + \frac{\eta}{h} \sum_{j=1}^m\mathbb{E} \left[  \mathbf{1}^t_i(j)\right]\mathbb{E} \left[w_i^t(j)\right]\mathbf{v}_{j} \nonumber\\
&=\mathbf{u}_i(t) + \frac{\eta}{h} \sum_{j=1}^mhp^{t}(\mathbf{s}_i^{\gamma}, \mathbf{v}_j; \alpha)(p_{pos}^t(\mathbf{u}_i, \mathbf{v}_{r_i^j}; \beta, \epsilon)-p_{neg}^t(\mathbf{u}_i, \mathbf{v}_{r_i^j}; \beta, \epsilon))\mathbf{v}_{j} \nonumber\\
&=\mathbf{u}_i(t) + \eta \sum_{j=1}^mp^{t}(\mathbf{s}_i^{\gamma}, \mathbf{v}_j; \alpha)g^{t}( \mathbf{u}_i, \mathbf{v}_j; \beta, \epsilon )\mathbf{v}_{j}.
\end{align}

\item First we have \(
\sum_{j=1}^m\mathbf{v}_j\mathbf{v}_j^T = \mathbf{V}\mathbf{V}^T,
\sum_{j=1}^m\mathbf{v}_j\sum_{k=1}^m\mathbf{v}_k^T = \mathbf{V}\mathbf{1}_{m \times m}\mathbf{V}^T
\), 
\begin{align}
&p^{t}(\mathbf{s}_i^{\gamma}, \mathbf{v}_j; \alpha) \nonumber\\
&= \frac{ \exp \left( \alpha \mathbf{v}_j^T \mathbf{s}_i^{\gamma}(t) \right) }{ \sum_{k = 1}^m \exp \left( \alpha \mathbf{v}_k^T \mathbf{s}_i^{\gamma}(t) \right) }= \frac{ 1 + \alpha \mathbf{v}_j^T \mathbf{s}_i^{\gamma}(t) +O\left( \alpha^2 (\mathbf{v}_j^T \mathbf{s}_i^{\gamma}(t))^2 \right) }{ \sum_{k = 1}^m  \left( 1+\alpha \mathbf{v}_k^T \mathbf{s}_i^{\gamma}(t) +O (\alpha^2 (\mathbf{v}_k^T \mathbf{s}_i^{\gamma}(t))^2 )\right) }\nonumber\\
&\approx \frac{ 1 + \alpha \mathbf{v}_j^T \mathbf{s}_i^{\gamma}(t)}{   m\left( 1+\frac{\alpha}{m} \sum_{k = 1}^m\mathbf{v}_k^T \mathbf{s}_i^{\gamma}(t)\right) }\approx \frac{1}{m}\left( 1 + \alpha \mathbf{v}_j^T \mathbf{s}_i^{\gamma}(t)-\frac{\alpha}{m} \sum_{k = 1}^m\mathbf{v}_k^T \mathbf{s}_i^{\gamma}(t)\right).\label{eq:similar-match-}
\end{align}
\begin{align}
g^{t}(\mathbf{u}_i, \mathbf{v}_j; \beta, \epsilon) &= \frac{ \left( 1 + \mathbf{v}_j^T \mathbf{u}_i(t) \right)^\beta - \left( 1 - \mathbf{v}_j^T \mathbf{u}_i(t) \right)^\beta }{ \left( 1 + \mathbf{v}_j^T \mathbf{u}_i(t) \right)^\beta + \left( 1 - \mathbf{v}_j^T \mathbf{u}_i(t) \right)^\beta } + \epsilon \nonumber\\
&\approx \frac{ \left( 1 + \beta\mathbf{v}_j^T \mathbf{u}_i(t) \right) - \left( 1 - \beta\mathbf{v}_j^T \mathbf{u}_i(t) \right) }{ \left( 1 + \beta\mathbf{v}_j^T \mathbf{u}_i(t) \right) + \left( 1 - \beta\mathbf{v}_j^T \mathbf{u}_i(t) \right) } + \epsilon \nonumber\\
&= \beta \mathbf{v}_j^T \mathbf{u}_i(t) + \epsilon.
\label{eq:confirm-bias}
\end{align}
Then Equation \ref{eq:pg} equals:
\begin{align}\label{eq:user-simp}
&\mathbf{u}_i(t+1)  \nonumber\\
&\approx \mathbf{u}_i(t) + \frac{\eta}{m} \sum_{j=1}^m ( 1 + \alpha \mathbf{v}_j^T \mathbf{s}_i^{\gamma}(t)-\frac{\alpha}{m} \sum_{k = 1}^m\mathbf{v}_k^T \mathbf{s}_i^{\gamma}(t))(\beta \mathbf{v}_j^T \mathbf{u}_i(t) + \epsilon)\mathbf{v}_{j} \nonumber\\
&\approx \mathbf{u}_i(t) + \frac{\eta}{m} \sum_{j=1}^m ( \alpha \epsilon \mathbf{v}_j^T \mathbf{s}_i^{\gamma}(t)-\frac{\alpha \epsilon}{m} \sum_{k = 1}^m\mathbf{v}_k^T \mathbf{s}_i^{\gamma}(t)+\beta \mathbf{v}_j^T \mathbf{u}_i(t) + \epsilon)\mathbf{v}_{j} \nonumber\\
&= \mathbf{u}_i(t) + \frac{\eta}{m} \sum_{j=1}^m ( (\alpha \epsilon \mathbf{v}_j^T -\frac{\alpha \epsilon}{m} \sum_{k = 1}^m\mathbf{v}_k^T )(\gamma \mathbf{u}_i(t) + (1 - \gamma) \frac{ \sum_{j \in \mathcal{N}_i} \mathbf{u}_j(t) }{ |\mathcal{N}_i| })\nonumber\\
&\ \ \ \ \ +\beta \mathbf{v}_j^T \mathbf{u}_i(t) + \epsilon)\mathbf{v}_{j} \nonumber\\
&=\frac{\eta \epsilon}{m}\sum_{j=1}^m \mathbf{v}_j+(\mathbf{I}_c+\frac{\eta(\alpha \epsilon \gamma+\beta)}{m}\sum_{j=1}^m\mathbf{v}_j\mathbf{v}_j^T-\frac{\eta \alpha \epsilon \gamma}{m^2}\sum_{j=1}^m\mathbf{v}_j\sum_{k=1}^m\mathbf{v}_k^T)\mathbf{u}_i(t)\nonumber\\
&\ \ \ \ \ +(\frac{\eta\alpha \epsilon (1-\gamma)}{m}\sum_{j=1}^m\mathbf{v}_j\mathbf{v}_j^T-\frac{\eta\alpha \epsilon (1-\gamma)}{m^2}\sum_{j=1}^m\mathbf{v}_j\sum_{k=1}^m\mathbf{v}_k^T)\frac{ \sum_{j \in \mathcal{N}_i} \mathbf{u}_j(t) }{ |\mathcal{N}_i| }\nonumber\\
&=\frac{\eta \epsilon}{m}\sum_{j=1}^m \mathbf{v}_j+(\mathbf{I}_c+\frac{\eta(\alpha \epsilon \gamma+\beta)}{m}\mathbf{V}\mathbf{V}^T-\frac{\eta \alpha \epsilon \gamma}{m^2}\mathbf{V}\mathbf{1}_{m \times m}\mathbf{V}^T)\mathbf{u}_i(t)\nonumber\\
&\ \ \ \ \ +(\frac{\eta\alpha \epsilon (1-\gamma)}{m}\mathbf{V}\mathbf{V}^T-\frac{\eta\alpha \epsilon (1-\gamma)}{m^2}\mathbf{V}\mathbf{1}_{m \times m}\mathbf{V}^T)\frac{ \sum_{j \in \mathcal{N}_i} \mathbf{u}_j(t) }{ |\mathcal{N}_i| }
\end{align} 
Under the definitions of $\mathbf{X}$, $\mathbf{Y}$, and $\mathbf{Z}$ given in Proposition \ref{thm:mat-format}, we have:
\begin{align}
&[\mathbf{u}_1(t+1),\mathbf{u}_2(t+1),\ldots,\mathbf{u}_n(t+1)]\nonumber\\&=\frac{\eta \epsilon}{m}[\sum_{j=1}^m\mathbf{v}_j, \sum_{j=1}^m\mathbf{v}_j, \ldots, \sum_{j=1}^m\mathbf{v}_j]\nonumber+\mathbf{Y}[\mathbf{u}_1(t),\mathbf{u}_2(t),\ldots,\mathbf{u}_n(t)]\\
&\ \ \ \ \ +\mathbf{Z}[\frac{ \sum_{j \in \mathcal{N}_1} \mathbf{u}_j(t) }{ |\mathcal{N}_1| }, \frac{ \sum_{j \in \mathcal{N}_2} \mathbf{u}_j(t) }{ |\mathcal{N}_2| }, \ldots, \frac{ \sum_{j \in \mathcal{N}_n} \mathbf{u}_j(t) }{ |\mathcal{N}_n| }] 
\end{align}
\begin{align}
\mathbf{U}(t+1)&=\frac{\eta \epsilon}{m}\mathbf{V}\mathbf{1}_{m \times n}+\mathbf{Y}\mathbf{U}(t)+\mathbf{Z}\mathbf{U}(t)(\operatorname{diag}(\mathbf{S}\mathbf{1}_{n \times 1})^{-1}\mathbf{S})^T \nonumber \\
&=\mathbf{X}+\mathbf{Y}\mathbf{U}(t)+ \mathbf{Z} \mathbf{U}(t) \mathbf{\tilde{S}}^T
\end{align}

\end{itemize}
\end{proof}

\section{Proof of Proposition \ref{thm:convergence}}
\label{apd:thm2}
\begin{proof}
\ 

Under the column-wise vectorization operator \(\ \operatorname{vec}(\cdot):\mathbb{R}^{c \times n} \rightarrow \mathbb{R}^{cn}\), Equation \ref{eq:update-mat-equiv} transforms to:
\begin{equation}
\operatorname{vec}(\mathbf{U}(t+1)) = \operatorname{vec}(\mathbf{X})+\operatorname{vec}(\mathbf{Y}\mathbf{U}(t))+\operatorname{vec}(\mathbf{Z} \mathbf{U}(t) \mathbf{\tilde{S}}^T).
\end{equation}
Based on Lemma \ref{lm:2}, we have 
\begin{align}
&\operatorname{vec}(\mathbf{Y}\mathbf{U}(t)) = \operatorname{vec}(\mathbf{Y}\mathbf{U}(t)\mathbf{I}^T_n) = (\mathbf{I}_n \otimes \mathbf{Y})\operatorname{vec}(\mathbf{U}(t))\ ;\\
&\operatorname{vec}(\mathbf{Z} \mathbf{U}(t) \mathbf{\tilde{S}}^T) = (\mathbf{\tilde{S}} \otimes \mathbf{Z})\operatorname{vec}(\mathbf{U}(t)).
\end{align}
Then
\begin{align}\label{eq:vec-func}
\operatorname{vec}(\mathbf{U}(t+1)) &= \operatorname{vec}(\mathbf{X}) +(\mathbf{I}_n \otimes \mathbf{Y})\operatorname{vec}(\mathbf{U}(t))+(\mathbf{\tilde{S}} \otimes \mathbf{Z})\operatorname{vec}(\mathbf{U}(t)) \nonumber \\
&= \operatorname{vec}(\mathbf{X}) +(\mathbf{I}_n \otimes \mathbf{Y}+\mathbf{\tilde{S}} \otimes \mathbf{Z})\operatorname{vec}(\mathbf{U}(t)).
\end{align}
Assume that $\mathbf{U}(t)$ in Equation \ref{eq:vec-func} converges to $\mathbf{U^*}$, we have 
\begin{equation}
(\mathbf{I}_{nc}-(\mathbf{I}_n \otimes \mathbf{Y}+\mathbf{\tilde{S}} \otimes \mathbf{Z}))\operatorname{vec}(\mathbf{U}^*) = \operatorname{vec}(\mathbf{X})
\end{equation}
Now, the convergence of Equation \ref{eq:update-mat-equiv} is equivalent to the invertibility of $(\mathbf{I}_{nc}-(\mathbf{I}_n \otimes \mathbf{Y}+\mathbf{\tilde{S}} \otimes \mathbf{Z}))$.
To establish that \(\eta(\frac{\beta}{2}+\frac{\alpha\epsilon\gamma}{2}+\frac{\beta^2}{8\alpha\epsilon\gamma}) < 1\) is sufficient for $(\mathbf{I}_{nc}-(\mathbf{I}_n \otimes \mathbf{Y}+\mathbf{\tilde{S}} \otimes \mathbf{Z}))^{-1}  exists$, we proceed through a sequence of logical implications:
\begin{enumerate}
\item We first demonstrate that \( \eta(\frac{\beta}{2}+\frac{\alpha\epsilon\gamma}{2}+\frac{\beta^2}{8\alpha\epsilon\gamma}) < 1 \Rightarrow \|\mathbf{Y}\|_\infty+ \|\mathbf{Z}\|_\infty < 1 \);\\
\textit{Proof.} We previously defined the representation of item $j$ as \(
\mathbf{v}_j = \left[ v_j^{(1)},v_j^{(2)}, \ldots, v_j^{(c)} \right]^T,j \in \{1,2, \ldots,m \}
\), if item $j$ belongs to category k, then
$v_j^{(k)}v_j^{(k)}=1,v_j^{(p)}v_j^{(q)}=0 \ \ for\ \  p \neq k \ \ or\ \  q\neq k$, and \(
\sum_{j=1}^mv_j^{(i)}=n_i,
\sum_{i=1}^cn_i=m
\).
\begin{align}\label{eq:VVT}
\mathbf{V}\mathbf{V}^T = \sum_{j=1}^m\mathbf{v}_j\mathbf{v}_j^T
&= 
\begin{bmatrix}
\sum_{j=1}^mv_j^{(1)}v_j^{(1)} & \cdots & \sum_{j=1}^mv_j^{(1)}v_j^{(c)} \\
\vdots & \ddots & \vdots \\
\sum_{j=1}^mv_j^{(c)}v_j^{(1)} & \cdots & \sum_{j=1}^mv_j^{(c)}v_j^{(c)}
\end{bmatrix} \nonumber\\
&=
\begin{bmatrix}
n_1 & 0   & \cdots & 0 \\
0   & n_2 & \cdots & 0 \\
\vdots & \vdots & \ddots & \vdots \\
0   & 0   & \cdots & n_c
\end{bmatrix}
\end{align}
\begin{align}
\mathbf{V}\mathbf{1}_{m \times m}\mathbf{V}^T &=\sum_{j=1}^m\mathbf{v}_j\sum_{k=1}^m\mathbf{v}_k^T \nonumber\\
&=\begin{bmatrix}
\sum_{j=1}^mv_j^{(1)}\sum_{k=1}^mv_k^{(1)} & \cdots & \sum_{j=1}^mv_j^{(1)}\sum_{k=1}^mv_k^{(c)} \\
\vdots & \ddots & \vdots \\
\sum_{j=1}^mv_j^{(c)}\sum_{k=1}^mv_k^{(1)} & \cdots & \sum_{j=1}^mv_j^{(c)}\sum_{k=1}^mv_k^{(c)}
\end{bmatrix} \nonumber \\
&=\begin{bmatrix}
n_1 \cdot n_1 & \cdots & n_1 \cdot n_c \\
\vdots & \ddots & \vdots \\
n_c \cdot n_1 & \cdots & n_c \cdot n_c
\end{bmatrix} 
\end{align}
\begin{align}
\mathbf{Y}&\approx \frac{\eta(\alpha \epsilon \gamma+\beta)}{m}
\begin{bmatrix}
n_1 & 0   & \cdots & 0 \\
0   & n_2 & \cdots & 0 \\
\vdots & \vdots & \ddots & \vdots \\
0   & 0   & \cdots & n_c
\end{bmatrix}-\frac{\eta \alpha \epsilon \gamma}{m^2}
\begin{bmatrix}
n_1 \cdot n_1 & \cdots & n_1 \cdot n_c \\
\vdots & \ddots & \vdots \\
n_c \cdot n_1 & \cdots & n_c \cdot n_c
\end{bmatrix} 
\end{align}

\begin{align}
\mathbf{Z}
&= \frac{\eta\alpha \epsilon (1-\gamma)}{m}(
\begin{bmatrix}
n_1 & 0   & \cdots & 0 \\
0   & n_2 & \cdots & 0 \\
\vdots & \vdots & \ddots & \vdots \\
0   & 0   & \cdots & n_c
\end{bmatrix}-\frac{1}{m}
\begin{bmatrix}
n_1 \cdot n_1 & \cdots & n_1 \cdot n_c \\
\vdots & \ddots & \vdots \\
n_c \cdot n_1 & \cdots & n_c \cdot n_c
\end{bmatrix}) 
\end{align}
\begin{align}
\|\mathbf{Y}\|_\infty&=\max_{1 \leq l \leq c}\left(\frac{\eta\alpha\epsilon\gamma}{m^2}n_l(\sum_{i=1}^cn_i)-\frac{\eta\alpha\epsilon\gamma}{m^2}n_l^2+\frac{\eta(\alpha\epsilon\gamma+\beta)}{m}n_l-\frac{\eta\alpha\epsilon\gamma}{m^2}n_l^2\right) \nonumber \\
&=\max_{1 \leq l \leq c}\left(\frac{\eta(2\alpha\epsilon\gamma+\beta)}{m}n_l-\frac{2\eta\alpha\epsilon\gamma}{m^2}n_l^2\right) \nonumber \\
&\leq\eta\frac{(2\alpha\epsilon\gamma+\beta)^2}{8\alpha\epsilon\gamma}
\end{align}

\begin{align}
\|\mathbf{Z}\|_\infty&=\frac{\eta\alpha \epsilon (1-\gamma)}{m}\max_{1 \leq l \leq c}\left(\frac{1}{m}n_l(\sum_{i=1}^cn_i)-\frac{1}{m}n_l^2+n_l-\frac{1}{m}n_l^2\right) \nonumber \\
&=\frac{2\eta\alpha \epsilon (1-\gamma)}{m}\max_{1 \leq l \leq c}\left(n_l-\frac{1}{m}n_l^2\right) \nonumber \\
&\leq \frac{2\eta\alpha \epsilon (1-\gamma)}{m} \cdot\frac{m}{4} \nonumber\\
&=\eta\frac{\alpha \epsilon (1-\gamma)}{2}
\end{align}
Then \(\|\mathbf{Y}\|_\infty+ \|\mathbf{Z}\|_\infty \leq \eta\frac{(2\alpha\epsilon\gamma+\beta)^2}{8\alpha\epsilon\gamma}+\eta\frac{\alpha \epsilon (1-\gamma)}{2}=\eta(\frac{\beta}{2}+\frac{\alpha\epsilon\gamma}{2}+\frac{\beta^2}{8\alpha\epsilon\gamma})< 1\). \hfill $\square$

\item Next, we demonstrate that \( \|\mathbf{Y}\|_\infty+ \|\mathbf{Z}\|_\infty < 1 \Rightarrow\|\mathbf{I}_n \otimes \mathbf{Y}+\mathbf{\tilde{S}} \otimes \mathbf{Z}\|_\infty < 1 \);\\
\textit{Proof.} Given the triangle inequality property of norms and Lemma \ref{lm:6}, we derive:
\[
\|\mathbf{I}_n \otimes \mathbf{Y}+\mathbf{\tilde{S}} \otimes \mathbf{Z}\|_\infty \leq \|\mathbf{I}_n \otimes \mathbf{Y}\|_\infty+\|\mathbf{\tilde{S}} \otimes \mathbf{Z}\|_\infty = \|\mathbf{I}_n\|_\infty \|\mathbf{Y}\|_\infty+\|\mathbf{\tilde{S}}\|_\infty \|\mathbf{Z}\|_\infty
\]
Furthermore, \(
\|\mathbf{I}_n\|_\infty =1\), and\ \(\mathbf{\tilde{S}}=\operatorname{diag}(\mathbf{S}\mathbf{1}_{n \times 1})^{-1}\mathbf{S}\), then \(\forall k,\sum_i|{\tilde{S}}_{ki}|=1,\|\mathbf{\tilde{S}}\|_\infty=1
\).
Consequently, we obtain
\(
\|\mathbf{I}_n \otimes \mathbf{Y}+\mathbf{\tilde{S}} \otimes \mathbf{Z}\|_\infty \leq \|\mathbf{Y}\|_\infty+ \|\mathbf{Z}\|_\infty < 1
\). \hfill $\square$
\item Finally, based on Lemma \ref{lm:5}, we have \( \|\mathbf{I}_n \otimes \mathbf{Y}+\mathbf{\tilde{S}} \otimes \mathbf{Z}\|_\infty < 1 \Rightarrow (\mathbf{I}_{nc}-(\mathbf{I}_n \otimes \mathbf{Y}+\mathbf{\tilde{S}} \otimes \mathbf{Z}))^{-1}  exists
\).
\end{enumerate}
Therefore, we prove that \(\eta(\frac{\beta}{2}+\frac{\alpha\epsilon\gamma}{2}+\frac{\beta^2}{8\alpha\epsilon\gamma}) < 1\) is sufficient for $(\mathbf{I}_{nc}-(\mathbf{I}_n \otimes \mathbf{Y}+\mathbf{\tilde{S}} \otimes \mathbf{Z}))^{-1}  exists$, i.e., the $\mathbf{U}(t)$ of Equation \ref{eq:update-mat-equiv} converges to \(\mathbf{U}^* = \operatorname{unvec}((\mathbf{I}_{nc}-(\mathbf{I}_n \otimes \mathbf{Y}+\mathbf{\tilde{S}} \otimes \mathbf{Z}))^{-1}\operatorname{vec}(\mathbf{X}))\).
\end{proof}

\section{Proof of COROLLARY \ref{thm:cluster}}
\label{apd:thm3}
\begin{proof}
\ 

\begin{itemize}[noitemsep,topsep=0pt,parsep=2pt,partopsep=0pt, leftmargin=1.5em]
\item Transient Homogenization \\
When \(
\epsilon = 0, \gamma=1
\), Equation \ref{eq:user-simp} reduces to 
\begin{equation}
\mathbf{u}_i(t+1)=(\mathbf{I}_c+\frac{\eta\beta}{m}\mathbf{V}\mathbf{V}^T)\mathbf{u}_i(t),
\end{equation}
Let \(\lambda=\frac{\eta\beta}{m}\), and applying Equation \ref{eq:VVT}, we have
\begin{align}\label{eq:user-up}
\begin{bmatrix}
u_i^{(1)}(t+1) \\
u_i^{(2)}(t+1) \\
\vdots \\
u_i^{(c)}(t+1)
\end{bmatrix} 
&= 
\begin{bmatrix}
1+\lambda n_1 & 0   & \cdots & 0 \\
0   & 1+\lambda n_2 & \cdots & 0 \\
\vdots & \vdots & \ddots & \vdots \\
0   & 0   & \cdots & 1+\lambda n_c
\end{bmatrix}
\begin{bmatrix}
u_i^{(1)}(t) \\
u_i^{(2)}(t) \\
\vdots \\
u_i^{(c)}(t)
\end{bmatrix} \nonumber\\
&=
\begin{bmatrix}
(1+\lambda n_1)u_i^{(1)}(t) \\
(1+\lambda n_2)u_i^{(2)}(t) \\
\vdots \\
(1+\lambda n_c)u_i^{(c)}(t)
\end{bmatrix}
\end{align}
If $\exists k$, s.t. \(
u_i^{(k)}(t)u_j^{(k)}(t) \geq \frac{\|\mathbf{u}_i(t)\|_2\|\mathbf{u}_j(t)\|_2}{\sqrt{1+(\frac{2n_k+\lambda n_k^2}{\sum_{l \neq k}2n_l+\lambda n_l^2})^2}}
\), based on Lemma \ref{lm:7}, we have 
\begin{align}
\min_{l \neq k}u_i^{(l)}(t)u_j^{(l)}(t) 
&\geq -\sqrt{\|\mathbf{u}_i(t)\|_2^2\|\mathbf{u}_j(t)\|_2^2-(\frac{\|\mathbf{u}_i(t)\|_2\|\mathbf{u}_j(t)\|_2}{\sqrt{1+(\frac{2n_k+\lambda n_k^2}{\sum_{l \neq k}2n_l+\lambda n_l^2})^2}})^2} \nonumber \\
&=-\frac{\|\mathbf{u}_i(t)\|_2\|\mathbf{u}_j(t)\|_2(2n_k+\lambda n_k^2)}{\sum_{l \neq k}(2n_l+\lambda n_l^2)\sqrt{1+(\frac{2n_k+\lambda n_k^2}{\sum_{l \neq k}2n_l+\lambda n_l^2})^2}}
\end{align}
Then 
\begin{align}
&\mathbf{u}_i^T(t+1)\mathbf{u}_j(t+1) - \mathbf{u}_i^T(t)\mathbf{u}_j(t) \nonumber\\
&=\sum_l(2\lambda n_l+\lambda^2n_l^2)u_i^{(l)}u_j^{(l)} \nonumber \\
&=\lambda \left((2 n_k+\lambda n_k^2)u_i^{(k)}u_j^{(k)} + \sum_{l \neq k}(2 n_l+\lambda n_l^2)u_i^{(l)}u_j^{(l)} \right) \nonumber \\
& \geq \lambda \left((2 n_k+\lambda n_k^2)u_i^{(k)}u_j^{(k)} + \sum_{l \neq k}(2 n_l+\lambda n_l^2)\min_{p \neq k}u_i^{(p)}u_j^{(p)} \right) \nonumber \\
&\geq \lambda \left(
\begin{aligned}
&(2 n_k+\lambda n_k^2)\frac{\|\mathbf{u}_i(t)\|_2\|\mathbf{u}_j(t)\|_2}{\sqrt{1+(\frac{2n_k+\lambda n_k^2}{\sum_{l \neq k}2n_l+\lambda n_l^2})^2}} \\
&- \sum_{l \neq k}(2 n_l+\lambda n_l^2)\frac{\|\mathbf{u}_i(t)\|_2\|\mathbf{u}_j(t)\|_2(2n_k+\lambda n_k^2)}{\sum_{l \neq k}(2n_l+\lambda n_l^2)\sqrt{1+(\frac{2n_k+\lambda n_k^2}{\sum_{l \neq k}2n_l+\lambda n_l^2})^2}} 
\end{aligned}
\right) \nonumber \\
&=0 
\end{align}

\item Steady-state Homogenization \\

\[
\text{Let }\mathbf{D}=\begin{bmatrix}
1+\lambda n_1 & 0   & \cdots & 0 \\
0   & 1+\lambda n_2 & \cdots & 0 \\
\vdots & \vdots & \ddots & \vdots \\
0   & 0   & \cdots & 1+\lambda n_c
\end{bmatrix},d_i = D_{ii} = 1+\lambda n_i
\]
\textbf{Proof by Mathematical Induction:}
\begin{equation}
\begin{aligned}
&\text{If } k = \operatorname*{argmax}_{o \in \{1,2,\ldots,c\}} n_o, \text{ then } \forall \tau \in \mathbb{Z}^{+}, \text{we have }\\
&u_i^{(k)}(t+\tau-1) u_j^{(k)}(t+\tau-1) \geq \frac{\|\mathbf{u}_i(t+\tau-1)\|_2 \|\mathbf{u}_j(t+\tau-1)\|_2}{\sqrt{1 + \left(\frac{2n_k + \lambda n_k^2}{\sum_{l \neq k} 2n_l + \lambda n_l^2}\right)^2}}.
\end{aligned}
\end{equation}
\textbf{Base Case:}  
We have verified that the inequality holds when we let \( \tau = 1 \), i.e.,
\begin{equation}
u_i^{(k)}(t)u_j^{(k)}(t) \geq \frac{\|\mathbf{u}_i(t)\|_2\|\mathbf{u}_j(t)\|_2}{\sqrt{1+(\frac{2n_k+\lambda n_k^2}{\sum_{l \neq k}2n_l+\lambda n_l^2})^2}},
\end{equation}
and we also have $d_k = 1 + \lambda n_k = 1 + \lambda \max_o n_o = \max_o(1+ \lambda n_o) = \max_o d_o$.

\textbf{Inductive Hypothesis:}  
Assume that the inequality holds for \( \tau = m \in \mathbb{Z}^{+}\), i.e.,
\begin{equation}
u_i^{(k)}(t+m-1)u_j^{(k)}(t+m-1) \geq \frac{\|\mathbf{u}_i(t+m-1)\|_2\|\mathbf{u}_j(t+m-1)\|_2}{\sqrt{1+(\frac{2n_k+\lambda n_k^2}{\sum_{l \neq k}2n_l+\lambda n_l^2})^2}}.
\end{equation}
\textbf{Inductive Step:}  
When \( \tau = m+1 \), based on Lemma \ref{lm:8}, we have
\begin{align}
&\frac{u_i^{(k)}(t+m)u_j^{(k)}(t+m)}{\|\mathbf{u}_i(t+m)\|_2\|\mathbf{u}_j(t+m)\|_2} \nonumber\\
&=\frac{d_k^2u_i^{(k)}(t+m-1)u_j^{(k)}(t+m-1)}{\|\mathbf{D}\mathbf{u}_i(t+m-1)\|_2\|\mathbf{D}\mathbf{u}_j(t+m)\|_2} \nonumber \\
& \geq \frac{d_k^2u_i^{(k)}(t+m-1)u_j^{(k)}(t+m-1)}{(\max_id_i)^2\|\mathbf{u}_i(t+m-1)\|_2\|\mathbf{u}_j(t+m-1)\|_2} \nonumber \\
&\geq \frac{d_k^2u_i^{(k)}(t+m-1)u_j^{(k)}(t+m-1)}{d_k^2\|\mathbf{u}_i(t+m-1)\|_2\|\mathbf{u}_j(t+m-1)\|_2} \nonumber \\
&= \frac{u_i^{(k)}(t+m-1)u_j^{(k)}(t+m-1)}{\|\mathbf{u}_i(t+m-1)\|_2\|\mathbf{u}_j(t+m-1)\|_2} \nonumber \\
& \geq \frac{1}{\sqrt{1+(\frac{2n_k+\lambda n_k^2}{\sum_{l \neq k}2n_l+\lambda n_l^2})^2}} 
\end{align}
\[
\text{i.e. }u_i^{(k)}(t+m)u_j^{(k)}(t+m) \geq \frac{\|\mathbf{u}_i(t+m)\|_2\|\mathbf{u}_j(t+m)\|_2}{\sqrt{1+(\frac{2n_k+\lambda n_k^2}{\sum_{l \neq k}2n_l+\lambda n_l^2})^2}}.
\]
After leveraging the inductive hypothesis, we show the inequality continues to hold when \( \tau = m+1 \).

\textbf{Conclusion:}  

By the principle of mathematical induction,

\begin{align}
&\forall \tau \in \mathbb{Z}^+ , \text{we have } \nonumber\\
&u_i^{(k)}(t+\tau-1)u_j^{(k)}(t+\tau-1) \geq\frac{\|\mathbf{u}_i(t+\tau-1)\|_2\|\mathbf{u}_j(t+\tau-1)\|_2}{\sqrt{1+(\frac{2n_k+\lambda n_k^2}{\sum_{l \neq k}2n_l+\lambda n_l^2})^2}},
\end{align}
\(
\text{so we also have }\forall \tau \in \mathbb{Z}^+ ,\mathbf{u}_i^T(t+\tau)\mathbf{u}_j(t+\tau) \geq \mathbf{u}_i^T(t+\tau-1)\mathbf{u}_j(t+\tau-1).
\)

\end{itemize}
\end{proof}

\section{Proof of COROLLARY \ref{thm:entropy}}
\label{apd:thm4}
\begin{proof}
\ 

From Equation \ref{eq:user-up}:
\(
\forall o,i,u_i^{(o)}(t+1)=(1+\lambda n_o)u_i^{(o)}(t)
\), and this is equivalent to 
\(
\Delta t=1, \frac{u_i^{(o)}(t+\Delta t)-u_i^{(o)}(t)}{\Delta t} =\lambda n_ou_i^{(o)}(t)
\), when 
\(
\Delta t \to 0\), we similarly obtain
\begin{equation}
\frac{\mathrm{d}u_i^{(o)}(t)}{\mathrm{d}t}=\frac{u_i^{(o)}(t+\Delta t)-u_i^{(o)}(t)}{\Delta t} =\lambda n_ou_i^{(o)}(t)
\end{equation}
Then
\begin{align}
p_{i,o}^t&:=\frac{s_i^o(t)}{\sum_{k=1}^cs_i^k(t)}=P(user\  i\  receives\  an\  item\  from\  Category\  o)\nonumber \\
&=\frac{n_oe^{\alpha u_i^{(o)}(t)}}{\sum_k n_ke^{\alpha u_i^{(k)}(t)}}
\end{align}
The time derivative of recommended categories entropy for user 
$i$ at time $t$ is given by
\begin{align}\label{eq:H}
&\frac{\mathrm{d}H_i(t)}{\mathrm{d}t} \nonumber \\
&=-\sum_{o=1}^c(\frac{\mathrm{d}p_{i,o}^t}{\mathrm{d}t}\ln p_{i,o}^t+p_{i,o}^t\frac{\mathrm{d}\ln p_{i,o}^t}{\mathrm{d}t}) 
=-\sum_{o=1}^c\ln p_{i,o}^t\frac{\mathrm{d}p_{i,o}^t}{\mathrm{d}t}-\sum_{o=1}^c\frac{\mathrm{d}p_{i,o}^t}{\mathrm{d}t} \nonumber \\
&=-\sum_{o=1}^c\ln p_{i,o}^t\frac{\mathrm{d}p_{i,o}^t}{\mathrm{d}t}-\frac{\mathrm{d}\sum_{o=1}^cp_{i,o}^t}{\mathrm{d}t}=-\sum_{o=1}^c\ln p_{i,o}^t\frac{\mathrm{d}p_{i,o}^t}{\mathrm{d}t}
\end{align}
And
\begin{align}
&\frac{\mathrm{d}p_{i,o}^t}{\mathrm{d}t} \nonumber\\
&= \frac{\frac{\mathrm{d}(n_oe^{\alpha u_i^{(o)}(t)})}{\mathrm{d}t}(\sum_k n_ke^{\alpha u_i^{(k)}(t)})-\frac{\mathrm{d}(\sum_k n_ke^{\alpha u_i^{(k)}(t)})}{\mathrm{d}t}(n_oe^{\alpha u_i^{(o)}(t)})}{(\sum_k n_ke^{\alpha u_i^{(k)}(t)})^2} \nonumber \\
&=\frac{(n_oe^{\alpha u_i^{(o)}(t)}\alpha\lambda n_ou_i^{(o)}(t))(\sum_k n_ke^{\alpha u_i^{(k)}(t)})}{(\sum_k n_ke^{\alpha u_i^{(k)}(t)})^2} \\
&\ \ \ \ -\frac{(\sum_k n_ke^{\alpha u_i^{(k)}(t)}\alpha \lambda n_ku_i^{(k)}(t))(n_oe^{\alpha u_i^{(o)}(t)})}{(\sum_k n_ke^{\alpha u_i^{(k)}(t)})^2}
\nonumber \\
&=\lambda \alpha \frac{n_oe^{\alpha u_i^{(o)}(t)}}{\sum_k n_ke^{\alpha u_i^{(k)}(t)}} \left(n_o u^{(o)}(t)-\sum_k n_k u_i^{(k)}(t)\frac{n_k e^{\alpha u_i^{(k)}(t)}}{\sum_kn_k e^{\alpha u_i^{(k)}(t)}} \right) \nonumber \\
&=\lambda \alpha p_{i,o}^t(n_ou_i^{(o)}(t)-\sum_kn_ku_i^{(k)}(t)p_{i,k}^t) 
\end{align}
Then Equation \ref{eq:H} equals:
\begin{align}\label{eq:XY}
&\frac{\mathrm{d}H_i(t)}{\mathrm{d}t} \nonumber \\
&=-\sum_{o=1}^c\ln p_{i,o}^t \lambda \alpha p_{i,o}^t(n_ou_i^{(o)}(t)-\sum_{k=1}^cn_ku_i^{(k)}(t)p_{i,k}^t) \nonumber \\
&=-\lambda \alpha \left(\sum_{o=1}^c(n_ou_i^{(o)}(t)\ln p_{i,o}^t)p_{i,o}^t-\sum_{k=1}^c(n_ku_i^{(k)}(t))p_{i,k}^t)(\sum_{o=1}^c (\ln p_{i,o}^t)p_{i,o}^t \right)
\end{align}
Let $X$ be a random variable with possible values \\
\(\{n_1u_i^{(1)}(t), n_2u_i^{(2)}(t),\ldots, n_cu_i^{(c)}(t)\}\), \\
let $Y$ be a random variable with possible values \\
\(\{\ln p_{i,1}^t, \ln p_{i,2}^t,\ldots, \ln p_{i,c}^t\}\), \\
and \(P(X=n_ou_i^{(o)}(t))=P(Y=\ln p_{i,o}^t)=p_{i,o}^t\).\\
Then Equation \ref{eq:XY} equals:
\begin{align}
\frac{\mathrm{d}H_i(t)}{\mathrm{d}t}&=-\lambda \alpha(\mathbb{E}[XY]-\mathbb{E}[X]\mathbb{E}[Y]) \nonumber \\
&=-\lambda \alpha \mathrm{Cov}(X,Y) 
\end{align}
Because \(p_{i,o}^t=\frac{n_oe^{\alpha u_i^{(o)}(t)}}{\sum_k n_ke^{\alpha u_i^{(k)}(t)}}\), an increase in $u_i^{(o)}(t)$ results in increases in both $n_ou_i^{(o)}(t)$ and $\ln p_{i,o}^t$, indicating a positive correlation between $X$ and $Y$, which implies $\mathrm{Cov}(X,Y)> 0$  and $\frac{\mathrm{d}H_i(t)}{\mathrm{d}t}<0$.
\end{proof}

\section{Supporting Lemmas}
\label{apd:lemma}
\begin{lemma}\label{lm:1}

\begin{equation}
\mathbb{E} \left[  \mathbf{1}^t_i(j)\right] \approx hp^{t}(\mathbf{s}_i^{\gamma}, \mathbf{v}_j; \alpha)
\end{equation}

\begin{proof}
\ \\
First we define the selection indicator:

\begin{equation}
\mathbf{1}^t_{i,k}(j) = 
\begin{cases} 
1, & \parbox[t]{0.75\linewidth}{
\raggedright if item $j$ is selected at the $k$\text{-}th draw\\
without replacement from $\mathbf{p}_{i}^{t}(\alpha)$
} \\
0, & \text{otherwise}
\end{cases}
\end{equation}

For simplicity, we denote $p^{t}(\mathbf{s}_i^{\gamma}, \mathbf{v}_j; \alpha)$ as $p_j$, because of $h \ll m, p_j \rightarrow 0$, we have:

\begin{equation}
\mathbb{E} \left[  \mathbf{1}^t_i(j)\right]=\sum_{k=1}^h\mathbb{E} \left[  \mathbf{1}^t_{i,k}(j)\right] \approx \sum_{k=1}^hp_j=hp_j
\end{equation}
Then we compute the error:
\begin{align}
\delta &= \sum_{k=1}^hp_j-\sum_{k=1}^h\mathbb{E} \left[  \mathbf{1}^t_{i,k}(j)\right] \approx hp_j-(p_j+(h-1)\sum_{l \neq j}p_lp_j) \nonumber\\
&= hp_j-(p_j+(h-1)(1-p_j)p_j) = (h-1)p_j^2 \nonumber\\
&= (h-1)(p^{t}(\mathbf{s}_i^{\gamma}, \mathbf{v}_j; \alpha))^2 
\end{align}

\end{proof}
\end{lemma}

\begin{lemma}\label{lm:2}
\begin{equation}
\operatorname{vec}(\mathbf{AU}\mathbf{B}^T) = (\mathbf{B} \otimes \mathbf{A})\operatorname{vec}(\mathbf{U})
\end{equation}
where \(
\mathbf{A} \in \mathbb{R}^{c \times c},\mathbf{U} \in \mathbb{R}^{c \times n}, \mathbf{B} \in \mathbb{R}^{n \times n}.
\)
\begin{proof}
\ \\
 \(\forall i,j,
(\mathbf{AU}\mathbf{B}^T)_{ij}=\sum_{k=1}^c\sum_{l=1}^nA_{ik}U_{kl}B^T_{lj}
\) is equal to the $(j-1)c+i$-th row element of matrix $\operatorname{vec}(\mathbf{AU}\mathbf{B}^T)$.

Furthermore, 
\begin{equation}
\mathbf{B} \otimes \mathbf{A} =
\begin{bmatrix}
B_{11} \mathbf{A} & \cdots & B_{1n} \mathbf{A} \\
\vdots & \ddots & \vdots \\
B_{n1} \mathbf{A} & \cdots & B_{nn} \mathbf{A}
\end{bmatrix} \in \mathbb{R}^{cn \times cn},
\end{equation}
the $(j-1)c+i$-th row element of matrix \((\mathbf{B} \otimes \mathbf{A})\operatorname{vec}(\mathbf{U})\) is equal to \(\sum_{k=1}^c\sum_{l=1}^nA_{ik}U_{kl}B^T_{lj}\).
Therefore, all corresponding elements of $\operatorname{vec}(\mathbf{AU}\mathbf{B}^T)$ and $(\mathbf{B} \otimes \mathbf{A})\operatorname{vec}(\mathbf{U})$ are equal, i.e., $\operatorname{vec}(\mathbf{AU}\mathbf{B}^T) = (\mathbf{B} \otimes \mathbf{A})\operatorname{vec}(\mathbf{U})$.
\end{proof}

\end{lemma}

\begin{lemma}\label{lm:3}
\begin{equation}
\|\mathbf{A}\mathbf{B}||_\infty \leq \|\mathbf{A}\|_\infty\|\mathbf{B}\|_\infty
\end{equation}

\begin{proof}
\ \\
For a matrix $\mathbf{A} \in \mathbb{R}_{m \times n}$, the infinity norm is defined as $\|\mathbf{A}\|_\infty=\max_{1 \leq i\leq m}\sum_{j=1}^n|A_{ij}|$.
Since \(
\sum_i|B_{ji}| \leq \max_{j}\sum_i|B_{ji}|=||\mathbf{B}||_\infty,\\
\sum_j|A_{kj}| \leq \max_{k}\sum_j|A_{kj}|=||\mathbf{A}||_\infty
\).\\
\begin{align}
\text{Then }\forall k, 
\sum_i|(\mathbf{A}\mathbf{B})_{ki}|&=\sum_i|\sum_jA_{kj}B_{ji}| \leq \sum_i \sum_j|A_{kj}||B_{ji}|\nonumber\\
&=\sum_j |A_{kj}|\sum_i|B_{ji}| \nonumber\\
&\leq \sum_j |A_{kj}|\|\mathbf{B}\|_\infty \leq \|\mathbf{A}\|_\infty \|\mathbf{B}\|_\infty
\end{align}

Therefore
\(
\|\mathbf{A}\mathbf{B}\|_\infty=\max_k \sum_i|(\mathbf{A}\mathbf{B})_{ki}| \leq \|\mathbf{A}\|_\infty \|\mathbf{B}\|_\infty
\).

\end{proof}
\end{lemma}

\begin{lemma}\label{lm:4}
\begin{equation}
\rho(\mathbf{A}) \leq \|\mathbf{A}\|_\infty
\end{equation}where $\rho(\mathbf{A})$ denotes the spectral radius of matrix $\mathbf{A}$.
\begin{proof}
\ \\
Let $\lambda$ be an eigenvalue of matrix $\mathbf{A}$, and let $\mathbf{x}$ be a corresponding eigenvector such that \(\mathbf{Ax}=\lambda\mathbf{x}\) and $\mathbf{x} \neq 0$.
Then 
\begin{equation}
\|\mathbf{Ax}\|_\infty=\|\lambda\mathbf{ x}\|_\infty = |\lambda| \|\mathbf{x}\|_\infty.
\end{equation}
Since \(\|\mathbf{Ax}\|_\infty \leq \|\mathbf{A}\|_\infty \|\mathbf{x}\|_\infty\) based on Lemma \ref{lm:3}, we get 
\begin{equation}
|\lambda| \|\mathbf{x}\|_\infty \leq \|\mathbf{A}\|_\infty \|\mathbf{x}\|_\infty.
\end{equation} 
And \(\|\mathbf{x}\|_\infty > 0\) with \(\mathbf{x} \neq 0\), we have \(|\lambda| \leq \|\mathbf{A}\|_\infty\), this inequality holds for all eigenvalues $\lambda$ of matrix $\mathbf{A}$. 
Therefore, 
\begin{equation}
\rho(\mathbf{A}) = \max \{ |\lambda| \big| \lambda \in \text{eigenvalues of } \mathbf{A} \} \leq \|\mathbf{A}\|_\infty.
\end{equation}

\end{proof}
\end{lemma}

\begin{lemma}\label{lm:5}
\ \\
A sufficient condition for the invertibility of 
 $\mathbf{I} - \mathbf{A}$ is that the infinity norm of $\mathbf{A}$ is strictly less than $1$, i.e.,
\begin{equation}
\|\mathbf{A}\|_\infty < 1 \Rightarrow (\mathbf{I} - \mathbf{A})^{-1} \text{ exists}.
\end{equation}
\begin{proof}
\ \\
Based on Lemma \ref{lm:4}, we get 
\(
\rho(\mathbf{A}) \leq \|\mathbf{A}\|_\infty.
\)
Given that $\|\mathbf{A}\|_\infty < 1$, it follows that
\begin{equation}
\rho(\mathbf{A}) < 1.
\end{equation}
This implies that all eigenvalues $\lambda$ of $\mathbf{A}$ satisfy $|\lambda| < 1$, i.e., all eigenvalues of $\mathbf{I} - \mathbf{A}$ satisfy
\begin{equation}
\mu = 1 - \lambda > 0.
\end{equation}
Hence, $\mathbf{I} - \mathbf{A}$ has no zero eigenvalues, which implies that
\begin{equation}
\det(\mathbf{I} - \mathbf{A}) \neq 0.
\end{equation}
Therefore, $\mathbf{I} - \mathbf{A}$ is invertible.
\end{proof}
\end{lemma}

\begin{lemma}\label{lm:6}
\begin{equation}
\|\mathbf{A} \otimes \mathbf{B}\|_\infty = \|\mathbf{A}\|_\infty\|\mathbf{B}\|_\infty
\end{equation}
\begin{proof}
\ \\
Since 
\begin{equation}
\|\mathbf{A}\|_\infty = \max_l \sum_i|A_{li}|, \|\mathbf{B}\|_\infty = \max_k \sum_r|A_{kr}|.
\end{equation}
Then 
\begin{align}
\|\mathbf{A} \otimes \mathbf{B}\|_\infty &=\max_{l,k}\sum_i \sum_r|A_{li}B_{kr}| \nonumber \\
&=\max_l \sum_i|A_{li}|\cdot \max_k \sum_r|A_{kr}| = \|\mathbf{A}\|_\infty\|\mathbf{B}\|_\infty
\end{align}
\end{proof}
\end{lemma}

\begin{lemma}\label{lm:7}

\[
\text{Let }
\mathbf{a} = [a^{(1)}, a^{(2)}, \ldots, a^{(n)}]^T \in \mathbb{R}^n, \  \mathbf{b} = [b^{(1)}, b^{(2)}, \ldots, b^{(n)}]^T \in \mathbb{R}^n.
\]
If there exists an index \(k \in \{1, \ldots, n\}\) such that \(a^{(k)} b^{(k)} \geq x > 0\), then a lower bound for the remaining products is given by
\begin{equation}
\min_{i \neq k} a^{(i)} b^{(i)} \geq -\sqrt{\|\mathbf{a}\|_2^2 \|\mathbf{b}\|_2^2 - x^2}.
\end{equation}
\begin{proof}
\ \\
First
\begin{align}
\sum_i(a^{(i)}b^{(i)})^2 &= \sum_i(a^{(i)})^2(b^{(i)})^2 \nonumber \\
&\leq \sum_i(a^{(i)})^2\sum_i(b^{(i)})^2 = \|\mathbf{a}\|_2^2\|\mathbf{b}\|_2^2
\end{align}
Given \(a^{(k)} b^{(k)} \geq x > 0\),
\begin{equation}
\sum_{i \neq k}(a^{(i)}b^{(i)})^2 \leq \|\mathbf{a}\|_2^2\|\mathbf{b}\|_2^2-x^2.
\end{equation}
This implies that for all \(i \neq k\),
\begin{equation}
|a^{(i)} b^{(i)}| \leq \sqrt{\|\mathbf{a}\|_2^2 \|\mathbf{b}\|_2^2 - x^2},
\end{equation}
which leads to:
\begin{equation}
\min_{i \neq k} a^{(i)} b^{(i)} \geq -\sqrt{\|\mathbf{a}\|_2^2 \|\mathbf{b}\|_2^2 - x^2}.
\end{equation}
\end{proof}
\end{lemma}

\begin{lemma}\label{lm:8}
\ \\
Let \( \mathbf{D} = \mathrm{diag}(d_1, d_2, \dots, d_n) \) be a diagonal matrix with strictly positive entries, i.e., \( d_i > 0 \) for all \( i \in \{1, 2, \dots, n\} \).  
Then, for any vector \( \mathbf{b} \in \mathbb{R}^n \), the following inequality holds:
\begin{equation}
\| \mathbf{D} \mathbf{b} \|_2 \leq \left( \max_i d_i \right) \cdot \| \mathbf{b} \|_2.
\end{equation}
\begin{proof}
\ \\
Let \( \mathbf{b} = [b_1, b_2, \ldots, b_n]^T \). Then,
\begin{equation}
\mathbf{D} \mathbf{b} = [d_1 b_1, d_2 b_2, \ldots, d_n b_n]^T,
\end{equation}
and 
\begin{align}
\| \mathbf{D} \mathbf{b} \|_2 &= \sqrt{\sum_{i=j}^n(d_jb_j)^2} \leq \sqrt{\sum_{j=1}^n( \max_i d_i)^2(b_j)^2} \nonumber \\
&=( \max_i d_i)\sqrt{\sum_{j=1}^n(b_j)^2} = ( \max_i d_i)\|\mathbf{b}\|_2
\end{align}
\end{proof}
\end{lemma}

